\documentclass[sigconf, screen]{acmart}
\renewcommand\footnotetextcopyrightpermission[1]{} %

\usepackage{fancyhdr}
\AtBeginDocument{%
  }

\usepackage[table,xcdraw,dvipsnames]{xcolor}
\usepackage{subcaption}
\usepackage{sansmath} %
\usepackage{soul}
\usepackage{xspace}
\usepackage{multirow}
\usepackage{xurl}
\usepackage{url}
\usepackage[many]{tcolorbox}
\usepackage{hyperref}
\usepackage{booktabs}
\usepackage{pifont}
\usepackage{float}
\usepackage[framemethod=TikZ]{mdframed}
\usepackage{tikz}
\usetikzlibrary{backgrounds, matrix,decorations.pathreplacing,calc,shapes,positioning,tikzmark,shadows.blur,automata, arrows.meta,fit}
\usepackage{wrapfig} 
\usepackage{graphicx}
\usepackage{stackengine}
\usepackage{makecell}
\usepackage{appendix}
\usepackage{adjustbox}
\usepackage[skip=\baselineskip]{caption}

\usepackage{changepage}  %
\usepackage{enumitem}[wide = 0pt, leftmargin=*]

\newfloat{lstfloat}{htbp}{lop}
\floatname{lstfloat}{Listing}

\newtcolorbox{thmbox}[1][]{colback=thm!5!white,colframe=thm!60!black,boxsep=-4pt,grow to left by=4pt,left=10pt,grow to right by=4pt,right=10pt,top=10pt,bottom=10pt,#1}
\newtcolorbox{defbox}[1][]{colback=def!5!white,colframe=def!60!black,boxsep=-4pt,grow to left by=4pt,left=10pt,grow to right by=4pt,right=10pt,top=10pt,bottom=10pt,#1}
\newtcolorbox{promptbox}[1][]{colback=prompt!5!white,colframe=prompt!60!black,boxsep=-4pt,grow to left by=4pt,left=10pt,grow to right by=4pt,right=10pt,top=10pt,bottom=10pt,#1}

\usepackage{algpseudocodex}
\usepackage{algorithm}
\usepackage{algorithmicx}

\usepackage{listings}

\usepackage{sankey} %

\newcommand{\smalltitle}[1]{\noindent\textbf{#1}}
\newcommand{\eg}{\textit{e.g.}} %
\newcommand{\ie}{\textit{i.e.}} %

\definecolor{mygray}{gray}{0.8}
\definecolor{mylightgray}{gray}{0.9}
\definecolor{mydarkgray}{gray}{0.3}

\definecolor{mygreen}{RGB}{200, 255, 200}
\definecolor{mydarkgreen}{rgb}{0,0.5,0}

\definecolor{myred}{RGB}{255, 200, 200}
\definecolor{mydarkred}{rgb}{0.5,0,0}

\definecolor{myblue}{RGB}{200, 200, 255}
\definecolor{mydarkblue}{rgb}{0,0,0.5}

\definecolor{mydarkdarkblue}{rgb}{0,0,0.7}
\definecolor{mydarkdarkgreen}{rgb}{0,0.7,0}
\definecolor{mydarkdarkgray}{gray}{0.1}

\definecolor{def}{RGB}{119, 228, 200}
\definecolor{thm}{RGB}{69, 53, 193}
\definecolor{prompt}{RGB}{255, 204, 153}

\newenvironment{packeditemize}{
\begin{list}{$\bullet$}{
\setlength{\itemsep}{1.5pt}
\setlength{\labelwidth}{8pt}
\setlength{\leftmargin}{10pt}
\setlength{\labelsep}{3pt}
\setlength{\listparindent}{\parindent}
\setlength{\parsep}{1.5pt}
\setlength{\parskip}{1.5pt}
\setlength{\topsep}{1.5pt}}}{\end{list}}

\newcommand{\work}[1]{\textsc{#1}\xspace}
\newcommand{\ins}[1]{\texttt{\MakeUppercase{#1}}}

\algrenewcommand\algorithmicrequire{\textbf{Input:}}
\algrenewcommand\algorithmicensure{\textbf{Output:}}

\AtBeginEnvironment{algorithmic}{\useupquotes}

\newcommand{\useupquotes}{%
  \begingroup\lccode`\~=`\'\lowercase{\endgroup\let~}\algoupquote
  \begingroup\lccode`\~=`\"\lowercase{\endgroup\let~}\algoupquotes
  \catcode`\'=\active\catcode`\"=\active
}

\newcommand{\algoupquote}{\mbox{\textquotesingle}}
\newcommand{\algoupquotes}{\mbox{\char`\"}}

\newtcbox{\mycolorbox}[1][red]
  {on line, arc = 0pt, outer arc = 0pt,
    colback = #1!10!white, colframe = #1!50!black,
    boxsep = 0pt, left = 1pt, right = 1pt, top = 2pt, bottom = 2pt,
    boxrule = 0pt, bottomrule = 1pt, toprule = 1pt}

\newtcbox{\rulebox}[1][green]
  {on line, arc = 0pt, outer arc = 0pt,
    colback = #1!10!white, colframe = #1!50!black,
    boxsep = 0pt, left = 1pt, right = 1pt, top = 1pt, bottom = 1pt,
    boxrule = 0pt, bottomrule = 0pt, toprule = 0pt}

\newtcbox{\llmbox}[1][red]
  {on line, arc = 0pt, outer arc = 0pt,
    colback = #1!10!white, colframe = #1!50!black,
    boxsep = 0pt, left = 1pt, right = 1pt, top = 1pt, bottom = 1pt,
    boxrule = 0pt, bottomrule = 0pt, toprule = 0pt}

\newtcbox{\insbox}[1][blue]
  {on line, arc = 0pt, outer arc = 0pt,
    colback = #1!10!white, colframe = #1!50!black,
    boxsep = 0pt, left = 1pt, right = 1pt, top = 1pt, bottom = 1pt,
    boxrule = 0pt, bottomrule = 0pt, toprule = 0pt}

\usepackage{circledsteps} %

\newcommand{\poc}{PoC\xspace}

\newcommand{\circlednumber}[2][white]{%
  \tikz[baseline=(char.base)]{
    \node[shape=circle, draw, fill=#1, minimum size=1pt, inner sep=0.5pt, text=black] (char) {#2};
  }%
}

\definecolor{codegreen}{rgb}{0,0.6,0}
\definecolor{clr-comment}{RGB}{0,128,0}

\definecolor{KeywordColor}{rgb}{0.0, 0.0, 0.6}    %
\definecolor{CommentColor}{rgb}{0.2, 0.5, 0.2}    %
\definecolor{StringColor}{rgb}{0.5, 0.0, 0.5}     %
\definecolor{BackgroundColor}{rgb}{0.98, 0.98, 0.98} %

\lstdefinelanguage{Solidity}{
	keywordstyle=\bfseries, %
	keywords=[1]{anonymous, assembly, assert, balance, break, call, callcode, case, catch, class, constant, continue, contract, debugger, default, delegatecall, delete, do, else, emit, event, experimental, export, external, false, finally, for, function, gas, if, implements, import, in, indexed, instanceof, interface, internal, is, length, library, log0, log1, log2, log3, log4, memory, modifier, new, payable, pragma, private, protected, public, pure, push, require, return, returns, revert, selfdestruct, send, solidity, storage, struct, suicide, super, switch, then, this, throw, transfer, true, try, typeof, using, view, while, with, addmod, ecrecover, keccak256, mulmod, ripemd160, sha256, sha3}, %
	keywordstyle=[1]\bfseries\color{KeywordColor},
	keywords=[2]{address, bool, byte, bytes, bytes1, bytes2, bytes3, bytes4, bytes5, bytes6, bytes7, bytes8, bytes9, bytes10, bytes11, bytes12, bytes13, bytes14, bytes15, bytes16, bytes17, bytes18, bytes19, bytes20, bytes21, bytes22, bytes23, bytes24, bytes25, bytes26, bytes27, bytes28, bytes29, bytes30, bytes31, bytes32, enum, int, int8, int16, int24, int32, int40, int48, int56, int64, int72, int80, int88, int96, int104, int112, int120, int128, int136, int144, int152, int160, int168, int176, int184, int192, int200, int208, int216, int224, int232, int240, int248, int256, mapping, string, uint, uint8, uint16, uint24, uint32, uint40, uint48, uint56, uint64, uint72, uint80, uint88, uint96, uint104, uint112, uint120, uint128, uint136, uint144, uint152, uint160, uint168, uint176, uint184, uint192, uint200, uint208, uint216, uint224, uint232, uint240, uint248, uint256, var, void, ether, finney, szabo, wei, days, hours, minutes, seconds, weeks, years},	%
	keywordstyle=[2]\bfseries\color{KeywordColor},
	keywords=[3]{block, blockhash, coinbase, difficulty, gaslimit, number, timestamp, msg, data, gas, sender, sig, now, tx, gasprice, origin},	%
	keywordstyle=[3]\bfseries\color{KeywordColor},
	keywords=[4]{constructor,makeAddr,startPrank,stopPrank,createSelectFork},	%
	keywordstyle=[4]\underline,
	keywords=[5]{MEM, MCOPY},	%
	keywordstyle=[5]\bfseries\color{red!80}\underline,
	keywords=[6]{v0, v1, v2, v3},	%
	keywordstyle=[6]\bfseries\color{orange!80}\underline,
	identifierstyle=,
	sensitive=true,
	comment=[l]{//},
	morecomment=[s]{/*}{*/},
	commentstyle=\color{gray}\itshape,
	stringstyle=\ttfamily\color{StringColor},
	morestring=[b]',
	morestring=[b]",
	morecomment=[f][\color{red}]{---}, 
	morecomment=[f][\color{codegreen}]{+++},
	morecomment=[f][\lstbg{red!20}]{-\ },
	morecomment=[f][\lstbg{green!20}]{+\ },
	morecomment=[f][\color{blue}]{@@},
	escapechar = \@
}

\lstdefinestyle{solidity_style}{
	language=Solidity,
	extendedchars=true,
	basicstyle=\linespread{0.8}\scriptsize\ttfamily,
    columns=fullflexible,
	showstringspaces=false,
	xleftmargin=1em,
	showspaces=false,
	numbers=left,
	numberstyle=\scriptsize\color{gray},
	numbersep=4pt,
	tabsize=2,
	frame=none,
	showtabs=true,
	captionpos=b,
	keepspaces=true,
	breaklines=true,
	escapeinside={/*!}{!*/}
}

\lstdefinelanguage{evm}{
	keywords=[1]{STOP, ADD, MUL, SUB, DIV, SDIV, MOD, SMOD, ADDMOD, MULMOD, EXP, SIGNEXTEND, LT, GT, SLT, SGT, EQ, ISZERO, AND, OR, XOR, NOT, BYTE, SHL, SHR, SAR, SHA3, ADDRESS, BALANCE, ORIGIN, CALLVALUE, CALLDATALOAD, CALLDATASIZE, CALLDATACOPY, CODESIZE, CODECOPY, GASPRICE, EXTCODESIZE, EXTCODECOPY, RETURNDATASIZE, RETURNDATACOPY, EXTCODEHASH, BLOCKHASH, COINBASE, TIMESTAMP, NUMBER, DIFFICULTY, GASLIMIT, CHAINID, SELFBALANCE, BASEFEE, POP, JUMP, JUMPI, PC, MSIZE, GAS, JUMPDEST, PUSH1, PUSH2, PUSH3, PUSH4, PUSH5, PUSH6, PUSH7, PUSH8, PUSH9, PUSH10, PUSH11, PUSH12, PUSH13, PUSH14, PUSH15, PUSH16, PUSH17, PUSH18, PUSH19, PUSH20, PUSH21, PUSH22, PUSH23, PUSH24, PUSH25, PUSH26, PUSH27, PUSH28, PUSH29, PUSH30, PUSH31, PUSH32, DUP1, DUP2, DUP3, DUP4, DUP5, DUP6, DUP7, DUP8, DUP9, DUP10, DUP11, DUP12, DUP13, DUP14, DUP15, DUP16, SWAP1, SWAP2, SWAP3, SWAP4, SWAP5, SWAP6, SWAP7, SWAP8, SWAP9, SWAP10, SWAP11, SWAP12, SWAP13, SWAP14, SWAP15, SWAP16, LOG0, LOG1, LOG2, LOG3, LOG4, RETURN, REVERT, INVALID, SELFDESTRUCT, NOP, CONST, LOG, THROW, THROWI, VAR, TXGASPRICE}, %
	keywordstyle=[1]\bfseries\color{KeywordColor},
	keywords=[2]{CALL, CALLCODE, DELEGATECALL, STATICCALL, CREATE, CREATE2},	%
	keywordstyle=[2]\bfseries\color{KeywordColor},
	keywords=[3]{SLOAD, SSTORE, MLOAD, MSTORE, MSTORE8},
	keywordstyle=[3]\bfseries\color{KeywordColor},
	identifierstyle=\color{black},
	sensitive=true,
	comment=[l]{//},
	morecomment=[s]{/*}{*/},
	commentstyle=\color{gray}\ttfamily,
	stringstyle=\color{red}\ttfamily,
	morestring=[b]',
	morestring=[b]"
}

\lstdefinestyle{evm_style}{
	language=evm,
	extendedchars=true,
	basicstyle=\linespread{0.8}\scriptsize\ttfamily,
    columns=fullflexible,
	showstringspaces=false,
	xleftmargin=2em,
	showspaces=false,
	numbers=left,
	numberstyle=\scriptsize\color{gray},
	numbersep=3pt,
	tabsize=2,
	frame=none,
	showtabs=true,
	captionpos=b,
	keepspaces=true,
	breaklines=true,
	morecomment=[f][\color{red}]{---}, 
	morecomment=[f][\color{codegreen}]{+++},
	morecomment=[f][\lstbg{red!20}]{-\ },
	morecomment=[f][\lstbg{green!20}]{+\ },
	morecomment=[f][\color{blue}]{@@},
	escapechar = \@
}

\newcommand{\sysname}{\textsc{TracExp}\xspace}
\newcommand{\dataset}{\textsc{TracExp-ds}\xspace}

\begin{document}

\title{From Transactions to Exploits: Automated PoC Synthesis for Real-World DeFi Attacks}

\begin{abstract}
Blockchain systems are increasingly targeted by on-chain attacks that exploit contract vulnerabilities to extract value rapidly and stealthily, making systematic analysis and reproduction highly challenging. In practice, reproducing such attacks requires manually crafting proofs-of-concept (PoCs), a labor-intensive process that demands substantial expertise and scales poorly.
In this work, we present the \textit{first} automated framework for synthesizing verifiable PoCs directly from on-chain attack executions. Our key insight is that attacker logic can be recovered from low-level transaction traces via trace-driven reverse engineering, and then translated into executable exploits by leveraging the code-generation capabilities of large language models (LLMs). To this end, we propose \sysname, which localizes attack-relevant execution contexts from noisy, multi-contract traces and introduces a novel dual-decompiler to transform concrete executions into semantically enriched exploit pseudocode. Guided by this representation, \sysname synthesizes PoCs and refines them to preserve exploitability-relevant semantics.
We evaluate \sysname on 321 real-world attacks over the past 20 months. \sysname successfully synthesizes PoCs for 93\% of incidents, with 58.78\% being directly verifiable, at an average cost of only \$0.07 per case. Moreover, \sysname enabled the release of a large number of previously unavailable PoCs to the community, earning a \$900 bounty and demonstrating strong practical impact.
\end{abstract}

\author{Xing Su$^{1}$, Hao Wu$^{1}$, Hanzhong Liang$^{1}$, Yunlin Jiang$^{1}$, Yuxi Cheng$^{1}$, Yating Liu$^{1}$, Fengyuan Xu$^{1,*}$}
\affiliation{
\institution{$^{1}$State Key Laboratory for Novel Software Technology, Nanjing University}
\country{China}
}

\email{xingsu@smail.nju.edu.cn, hao.wu@nju.edu.cn}
\email{{hanz_liang,231220040,yuxicheng,yatingliu}@smail.nju.edu.cn}
\email{fengyuan.xu@nju.edu.cn}
\thanks{*Corresponding author.}

\maketitle

\section{Introduction}\label{sec:introduction}

The decentralized finance (DeFi) ecosystem has experienced explosive growth in recent years, enabling complex financial primitives such as flash loans, decentralized exchanges, and permissionless lending protocols. As of January 2026, DeFi protocols collectively manage nearly \$130 billion in total value locked (TVL)~\cite{defillama}. At the core of this prosperity are smart contracts, autonomous programs that directly control and transfer substantial on-chain assets without intermediaries.
While this design enables transparency and composability, it also amplifies security risks: any flaw in business logic design or implementation can be exploited with scalable losses~\cite{AttackSoK,sok_rootcause}. In 2025 alone, on-chain attacks caused losses exceeding \$2.9 billion~\cite{SlowMist_Team}, with many incidents involving complex interactions across multiple contracts and transactions.

Effectively responding to such incidents requires more than detecting attacks~\cite{txspector,clue}, it critically depends on \emph{understanding how exploits were carried out} in sufficient detail to support incident response, root-cause analysis, and the design of effective defenses.
In practice, this level of understanding is most reliably achieved through  proof-of-concept (PoC), which provide a precise, verifiable, and end-to-end representation of attack behavior.
While manually curated PoCs remain labor-intensive and require expert knowledge, community efforts such as \work{DeFiHackLabs}~\cite{DeFiHackLabs} have made significant contributions by collecting hundreds of accessible PoCs written by security experts for real-world attacks. 
Nevertheless, our measurement of wild incidents indicates that $\sim$50\% of attacks still lack PoCs, and as the number and complexity of incidents continue to grow, this gap is likely to widen, highlighting the urgent need for scalable and automated approaches.

Fortunately, recent developments provide a practical opportunity to achieve this automation. On-chain transactions record \textit{concrete, path-sensitive} executions of real-world exploits, capturing the exact information (\eg, fund flows) during an attack. Moreover, decades of progress in EVM reverse engineering have yielded decompilers~\cite{Phalcon,Openchain,tenderly,dedaub,skylens} that lift low-level bytecode into structured, human-readable representations, offering a potential bridge from raw traces to higher-level program logic. In parallel, recent advances in LLMs have demonstrated strong capabilities in code generation~\cite{code_generation_survey,code_generation_survey2}. Together, these developments suggest a promising direction: \textit{automatically synthesizing PoCs by transforming execution evidence into concise and reproducible exploit logic}.

However, turning this opportunity into a practical system poses non-trivial challenges.
Execution traces are low-level, verbose, and span multiple contracts (\eg, attacker contracts, victim protocols, and token contracts), making it difficult to isolate attack-relevant logic directly.
While existing decompilers offer structured views of EVM bytecode, they are fundamentally designed for whole-program analysis and decoupled from transaction-level execution context. Besides, they struggle to faithfully capture dynamic behaviors critical to exploits, such as call arguments, data flows, and fund flows~\cite{smartcat,ObfProbe,skanf}.
Finally, synthesizing \emph{verifiable} PoCs is inherently attack-specific. Although LLMs can generate source code, their hallucinations~\cite{code_hallucination_1,code_hallucination_2} prevent guarantees that the PoCs are reproduce the original exploit, making validation essential.

To address these challenges, we propose \sysname, the \textit{first} automated framework for synthesizing verifiable PoCs directly from on-chain attack transactions.
\sysname first localizes attack-relevant execution contexts from noisy, multi-contract traces and lifts them into a compact, contract-centric representation.
It then introduces a \emph{trace-driven dual decompiler} that integrates static decompilation with dynamic values observed during execution, enabling deterministic recovery of high-level exploit pseudocode from a single concrete trace.
Guided by this representation, \sysname leverages LLMs for PoC synthesis and employs an \emph{exploit-aware refinement loop} that validates reproduced exploits via fund-flow oracles, prioritizing semantic exploitability over strict trace equivalence.

We evaluate \sysname on 321 real-world DeFi attacks collected over 20 months.
\sysname successfully synthesizes PoCs for 93\% of these incidents, with roughly half being directly verifiable, at an average runtime $<$5 minutes and monetary cost of $\sim$\$0.07 per case.
Beyond controlled evaluation, \sysname enabled the contribution of 33 previously unavailable PoCs to the community within 2 days, accounting for 38\% of all submissions during that month (ranked 1st) and earning \$900 bounty.
Measured per attack event, this output substantially exceeds that of experts, demonstrating markedly higher efficiency and effectiveness than manual PoC construction.

In summary, we make the following contributions:

\begin{packeditemize}
\item {\bf New Framework.} 
We propose \sysname, the \textit{first} general framework that automatically synthesizes verifiable \poc directly from on-chain attack transactions.
\sysname is attack-agnostic and requires no prior knowledge of vulnerability types or source code, enabling robust reproduction across diverse real-world attacks.

\item {\bf New Method.} 
We introduce a \textit{trace-driven reverse engineering} technique that integrates static decompilation with dynamic values from attack traces, enabling a \textit{dual decompiler} to deterministically recover concise, high-level exploit pseudocode from a single execution, overcoming imprecision that limit static analysis.

\item {\bf New Validation Mechanism.} 
We design an \textit{exploit-aware validation and refinement mechanism} that leverages fund-flow oracles to ensure semantic exploitability, while automatically repairing syntactic and semantic deviations in synthesized PoCs, guaranteeing they are both compilable and faithful to the original exploit.

\item {\bf Large-Scale Evaluation.} 
We conduct one of the \textit{largest} real-world on-chain attack reproductions.
\sysname successfully synthesizes PoCs for 93\% of evaluated incidents at an average monetary cost of \$0.07 per case, with 58.78\% of them being directly verifiable.

\item {\bf Real-World Impact.}
We demonstrate the practical utility by contributing 33 previously unavailable PoCs to the community within 2 days, showing that automated PoC synthesis can surpass expert-driven efforts in both efficiency and effectiveness.

\end{packeditemize}

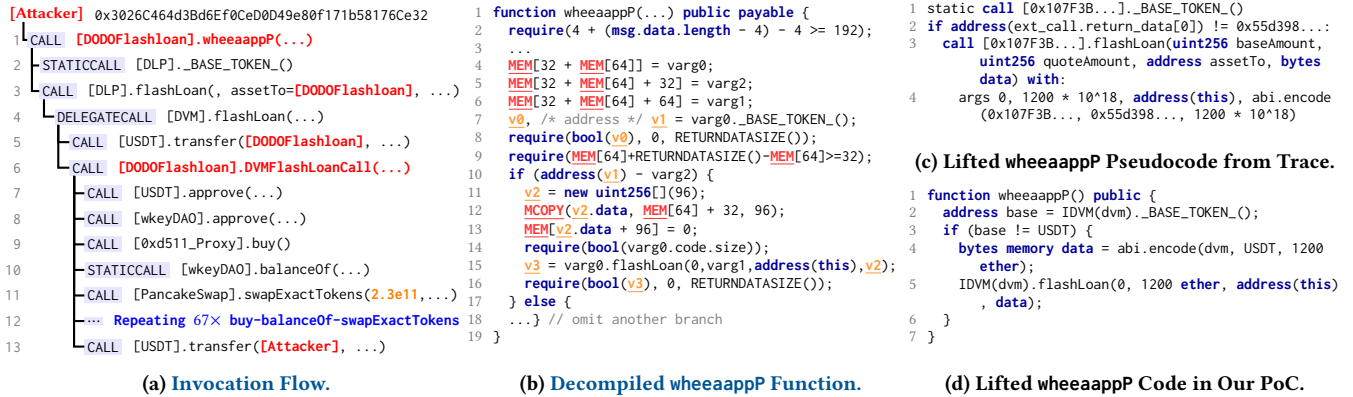
\begin{figure*}[t]
\begin{subfigure}[b]{0.35\textwidth}
\begin{tikzpicture}[x=0.2cm,y=0.34cm,
    attacker/.style={anchor=west, font=\sffamily\scriptsize, align=center, inner sep=0pt},
    calltype/.style={anchor=west, font=\sffamily\scriptsize, align=center, inner sep=1pt, fill=blue!10},
    nodetext/.style={anchor=west, font=\ttfamily\scriptsize, align=center, inner sep=1pt},
    toggle/.style={anchor=west, draw=black, fill=white, inner sep=0pt, font=\tiny, minimum size=3pt},
    coord/.style={font=\scriptsize, align=center, inner sep=0pt},
    lineno/.style={anchor=east, font=\ttfamily\scriptsize, align=center, inner sep=1pt, text=gray},
    funcbox/.style={draw=red, thick, rounded corners, inner sep=2pt, fill=yellow!10},
    codebox/.style={fill=gray!5, inner sep=0mm, text width=6.5cm, align=left}
]
\node[attacker] at (0, 0) (attacker) {\textcolor{red}{\textbf{[Attacker]}}};
\node[nodetext, right=1mm of attacker.east] {0x3026C464d3Bd6Ef0CeD0D49e80f171b58176Ce32};
\node[calltype] at (1.2, -1) (1call) {$\texttt{CALL}$};
\node[nodetext, right=1mm of 1call.east] (1callt) {\textcolor{red}{\textbf{[DODOFlashloan].wheeaappP(...)}}};
\node[calltype] at (2, -2) (2call) {$\texttt{STATICCALL}$};
\node[nodetext, right=1mm of 2call.east] {[DLP].\_BASE\_TOKEN\_()};
\node[calltype] at (2, -3) (3call) {$\texttt{CALL}$};
\node[nodetext, right=1mm of 3call.east] {[DLP].flashLoan(, assetTo=\textcolor{red}{\textbf{[DODOFlashloan]}}, ...)};
\node[calltype] at (3, -4) (4call) {$\texttt{DELEGATECALL}$};
\node[nodetext, right=1mm of 4call.east] {[DVM].flashLoan(...)};
\node[calltype] at (4, -5) (5call) {$\texttt{CALL}$};
\node[nodetext, right=1mm of 5call.east] {[USDT].transfer(\textcolor{red}{\textbf{[DODOFlashloan]}}, ...)};
\node[calltype] at (4, -6) (6call) {$\texttt{CALL}$};
\node[nodetext, right=1mm of 6call.east] {\textcolor{red}{\textbf{[DODOFlashloan].DVMFlashLoanCall(...)}}};
\node[calltype] at (5, -7) (7call) {$\texttt{CALL}$};
\node[nodetext, right=1mm of 7call.east] {[USDT].approve(...)};
\node[calltype] at (5, -8) (8call) {$\texttt{CALL}$};
\node[nodetext, right=1mm of 8call.east] {[wkeyDAO].approve(...)};
\node[calltype] at (5, -9) (9call) {$\texttt{CALL}$};
\node[nodetext, right=1mm of 9call.east] (9callt) {[0xd511\_Proxy].buy()};
\node[calltype] at (5, -10) (10call) {$\texttt{STATICCALL}$};
\node[nodetext, right=1mm of 10call.east] (10callt) {[wkeyDAO].balanceOf(...)};
\node[calltype] at (5, -11) (11call) {$\texttt{CALL}$};
\node[nodetext, right=1mm of 11call.east] (11callt) {[PancakeSwap].swapExactTokens(\textcolor{orange}{\textbf{2.3e11}},...)};
\node[calltype] at (5, -12) (12call) {$...$};
\node[nodetext, right=1mm of 12call.east] {\textcolor{blue}{\textbf{Repeating $67 \times$ buy-balanceOf-swapExactTokens}}};
\node[calltype] at (5, -13) (13call) {$\texttt{CALL}$};
\node[nodetext, right=1mm of 13call.east] {[USDT].transfer(\textcolor{red}{\textbf{[Attacker]}}, ...)};

\foreach \i in {1,...,13} {
    \node[lineno] at ($(0, 0)+(2mm, -\i*3.4mm)$) {\i};
}

\draw[-, thick] (attacker.198) |- coordinate(pos) (1call);
\draw[-, thick] (1call.210) |- (2call);
\draw[-, thick] (1call.210) |- (3call);
\draw[-, thick] (3call.210) |- (4call);
\draw[-, thick] (4call.190) |- (5call);
\draw[-, thick] (4call.190) |- (6call);
\draw[-, thick] (6call.210) |- (7call);
\draw[-, thick] (6call.210) |- (8call);
\draw[-, thick] (6call.210) |- (9call);
\draw[-, thick] (6call.210) |- (10call);
\draw[-, thick] (6call.210) |- (11call);
\draw[-, thick] (6call.210) |- (12call);
\draw[-, thick] (6call.210) |- (13call);

\end{tikzpicture}
\caption{\href{https://app.blocksec.com/explorer/tx/bsc/0xc9bccafdb0cd977556d1f88ac39bf8b455c0275ac1dd4b51d75950fb58bad4c8}{Invocation Flow.}}
\label{fig:invocation_flow}
\end{subfigure}
\begin{subfigure}[b]{0.32\textwidth}
\begin{lstlisting}[style=solidity_style,linewidth=\columnwidth]
function wheeaappP(...) public payable {
  require(4 + (msg.data.length - 4) - 4 >= 192);
  ...
  MEM[32 + MEM[64]] = varg0; /*!\label{dec_line:4}!*/
  MEM[32 + MEM[64] + 32] = varg2;
  MEM[32 + MEM[64] + 64] = varg1; /*!\label{dec_line:6}!*/
  v0, /* address */ v1 = varg0._BASE_TOKEN_();
  require(bool(v0), 0, RETURNDATASIZE()); 
  require(MEM[64]+RETURNDATASIZE()-MEM[64]>=32);
  if (address(v1) - varg2) {
    v2 = new uint256[](96);
    MCOPY(v2.data, MEM[64] + 32, 96);
    MEM[v2.data + 96] = 0;
    require(bool(varg0.code.size));
    v3 = varg0.flashLoan(0,varg1,address(this),v2); /*!\label{dec_line:15}!*/
    require(bool(v3), 0, RETURNDATASIZE()); 
  } else {
  ...} // omit another branch /*!\label{dec_line:17}!*/ 
}
\end{lstlisting}
\caption{\href{https://app.dedaub.com/binance/address/0x3783c91ee49a303c17c558f92bf8d6395d2f76e3/decompiled}{Decompiled \texttt{wheeaappP} Function.}}
\label{fig:decompiled_attack_functions}
\end{subfigure}
\begin{minipage}[b]{0.32\textwidth}
\begin{subfigure}[b]{\linewidth} 
\begin{lstlisting}[style=solidity_style,linewidth=\columnwidth, breaklines=true]
static call [0x107F3B...]._BASE_TOKEN_()
if address(ext_call.return_data[0]) != 0x55d398...:
  call [0x107F3B...].flashLoan(uint256 baseAmount, uint256 quoteAmount, address assetTo, bytes data) with:
    args 0, 1200 * 10^18, address(this), abi.encode(0x107F3B..., 0x55d398..., 1200 * 10^18)
\end{lstlisting}
\caption{Lifted \texttt{wheeaappP} Pseudocode from Trace.}
\label{fig:dec_attack_functions_in_trace}
\end{subfigure}
\begin{subfigure}[b]{\linewidth} 
\begin{lstlisting}[style=solidity_style,linewidth=\columnwidth, breaklines=true]
function wheeaappP() public {
  address base = IDVM(dvm)._BASE_TOKEN_();
  if (base != USDT) {
    bytes memory data = abi.encode(dvm, USDT, 1200 ether);
    IDVM(dvm).flashLoan(0, 1200 ether, address(this), data);
  }
}
\end{lstlisting}
\caption{Lifted \texttt{wheeaappP} Code in Our PoC.}
\label{fig:attack_functions_in_poc}
\end{subfigure}
\end{minipage}
\caption{Running example from the WebkeyDAO attack~\cite{WebKeyDaoIncident}. 
(a) involves >15 contracts, $\sim$5K cross-contract calls, and 132 million low-level EVM instructions.
(b) shows decompiled \texttt{wheeaappP} code produced by SOTA decompiler~\cite{gigahorse,elipmoc}, exposing numerous unoptimized memory-related operations (\textcolor{red!80}{red color}).
(c) presents our trace-driven lifted pseudocode.
(d) shows the lifted \texttt{wheeaappP} function in the PoC, which captures the core attack logic with explicit semantics (\eg, \texttt{address(this)}), lifted unknown call targets \texttt{dvm} and parameters \texttt{data}, while eliminating low-level memory artifacts.
}
\label{fig:running_example}
\Description{Running Example: WebkeyDAO attack~\cite{WebKeyDaoIncident}.}
\end{figure*}

\section{Background and Motivation}\label{sec:background}

In this section, we first provide some background knowledge on blockchain and DeFi attacks (\S~\ref{sec:background:defi_attack}), and then we introduce existing DeFi attacks analysis techniques and discuss their limitations (\S~\ref{sec:background:analysis}).

\subsection{A Primer on Blockchain and DeFi Attacks}\label{sec:background:defi_attack}

\smalltitle{Blockchain and Smart Contracts.}
Blockchains are decentralized, tamper-resistant ledgers recording \textit{transactions}. Platforms like Ethereum and BNB Smart Chain extend this model with \textit{smart contracts}, self-executing programs that automate trustless interactions. Transactions transfer value, invoke contract functions, or deploy code, and originate from either \textit{externally owned accounts} (EOAs) or \textit{contract accounts}. \textit{Tokens}, standardized via protocols like ERC-20~\cite{erc20} and ERC-721~\cite{erc721}, and stablecoins (\eg, USDT~\cite{usdt}) provide liquidity and underpin most on-chain financial activities.

\smalltitle{DeFi Protocols.}
DeFi protocols leverage smart contracts to offer permissionless financial services such as lending (\eg, Aave~\cite{aave}) and asset management. Two key primitives underpin these systems:
(1) \textit{Token swaps}, typically implemented by AMMs, enable decentralized trading against liquidity pools; 
(2) \textit{Flash loans} (\eg, Uniswap~\cite{uniswap}) allow users to borrow assets without collateral within a single transaction. 
Composability (\ie, the integration of these primitives) enables complex financial workflows but also amplifies risk.

\smalltitle{DeFi Attacks.}
The atomic and composable nature of DeFi can be exploited for complex attacks. In 2025, the incidents caused $>$\$2.9B in losses~\cite{SlowMist_Team}. Common attack vectors include reentrancy, price manipulation, and flash loan attacks. The attacker typically deploy \textit{adversary contracts} to orchestrate the exploit, coordinating interactions with victim protocols and auxiliary services (\eg, DEXs), and our preliminary analysis of 300+ incidents shows that \textbf{$>$90\% of attacks involve adversary contracts}. 

\textit{Example: WebkeyDAO Attack~\cite{WebKeyDaoIncident}.} 
\autoref{fig:invocation_flow} shows a partial invocation flow of a real-world WebkeyDAO attack~\cite{WebKeyDaoIncident}. The attacker (0x3026C46…) deployed an adversary contract (DODOFlashloan) and triggered the exploit via the \texttt{wheeaappP} function. During execution, the adversary contract coordinated interactions with the other contracts (\eg, \texttt{wkeyDAO}) and executed callbacks (Line~6), performing buy/sell operations (Lines 9–12) that ultimately transferred assets, yielding $\sim$\$737,000 in profit.

\subsection{Existing Techniques and Limitations}\label{sec:background:analysis}

To analyze on-chain attacks, several techniques have been proposed, ranging from low-level execution tracing to static decompilation and manual PoC construction. 
While each approach provides valuable insights, they face distinct limitations in scalability, automation, and semantic precision. 

\smalltitle{Tracing.}
Tracing is a fundamental technique for analyzing on-chain transactions, as it records all executed low-level instructions (\ie, execution traces) across contracts. In the running example, the full instruction-level trace contains $>$132 million EVM instructions. To improve usability, Ethereum clients (\eg, Geth~\cite{geth}) provide call tracers that retain only call-related events, reducing the trace to $\sim$5,000 instructions. Commercial tools~\cite{Phalcon,Openchain,tenderly} further enrich call traces with external knowledge (\eg, ABIs, and name tags) to facilitate understanding (\eg, the contract name \texttt{DLP} in Line~2). 

However, even call-level traces remain difficult to analyze in practice. They still involve complex cross-contract interactions (15 contracts and $\sim$5K calls in our example) and lack explicit data-flow semantics(\eg, the value \texttt{2.3e11} in Line~11 corresponds to the return value of a preceding call). Moreover, folded execution structures (\eg, loops in Line~12) can substantially inflate trace length, obscuring the underlying attack logic. Our empirical study of $>$300 incidents confirms this complexity: attack traces contain on average $>$770K executed instructions, involve 17 contracts, and include $>$1,700 cross-contract calls. 
As a result, tracing alone provides rich but low-level evidence, requiring labor-intensive manual inspection to reconstruct the underlying exploit logic.

\smalltitle{EVM Decompilation.}
Since most on-chain contracts ($\sim$99\%) are not open-sourced (\ie, verified)~\cite{etherscan}, another common approach is to decompile attacker bytecode into human-readable code~\cite{gigahorse,elipmoc,disco,Panoramix,Heimdall,Shrnkr,evm_decompiler}.

Despite these efforts, decompiled outputs are often insufficient for precise recovery of attack logic. First, call parameters and intermediate values may remain unresolved~\cite{smartcat} (\eg, variable \texttt{v2} in Line~\ref{dec_line:15}), because static decompilers are largely path-insensitive (Lines~\ref{dec_line:4}–\ref{dec_line:15}) and conservatively merge multiple execution paths. Consequently, memory states specific to feasible attack paths cannot be reconstructed precisely. 
More fundamentally, even along a single feasible path, EVM-specific dynamics (\eg, unknown loop bounds, runtime-dependent parameters, and call return values) hinder accurate memory modeling. We model EVM memory as
\begin{equation*}
\setlength{\abovedisplayskip}{3pt}
\setlength{\belowdisplayskip}{3pt}
D_M : (\mathit{range}, \mathit{len}) \mapsto e,
\end{equation*}
where both memory regions $(\mathit{range}, \mathit{len})$ and semantic expressions $e$ may remain symbolic, causing frequent aliasing and preventing reliable elimination of low-level artifacts (\eg, \texttt{MEM}). Finally, because static decompilation over-approximates all possible branches, isolating the core attack logic (\eg, Line~\ref{dec_line:17}) becomes difficult, especially when adversaries deliberately obfuscate control flow or distribute logic across contracts. Our evaluation confirms that even advanced decompilers retain substantial low-level artifacts, limiting their usefulness for exploit reconstruction.

\smalltitle{PoC Generation.}
Constructing proof-of-concept (PoC) exploits is a widely adopted strategy for validating and understanding attacks. Frameworks such as \work{Foundry}~\cite{foundry} support this process by enabling historical state replay and account impersonation. Community efforts like \work{DeFiHackLabs}~\cite{DeFiHackLabs} have curated over 670 PoCs, providing valuable reference implementations.
Nevertheless, PoC construction remains largely manual and labor-intensive. Our measurement shows that $\sim$50\% of real-world incidents still lack corresponding PoCs (see~\S~\ref{sec:evaluation}), creating a significant coverage gap.

In summary, while tracing, decompilation, and manual \poc construction have become indispensable for analyzing on-chain attacks, their limitations in scalability, automation, and semantic accuracy prevent comprehensive coverage of real-world exploits. This motivates our work: \textit{we propose to harness the code generation capabilities of LLMs to automate \poc construction, bridging the gap between low-level execution traces and high-level, reproducible exploit logic}. 
\section{Overview}\label{sec:overview}

\subsection{Problem Formalization}

\smalltitle{Goal.}
We aim to automatically synthesize proof-of-concept (\poc) at the \textit{source-code level} from observed attack transactions.
Given the execution traces $T=\langle i_1, ..., i_n\rangle$ during an attack, \sysname lifts these low-level instructions $i$ into a source-level exploit program $P$ that encodes the adversarial execution logic.
The synthesized program $P$ is considered correct if, when executed under a compatible on-chain state, it reproduces the execution behavior and unintended asset (\eg, USDT in our example) transfers observed in the original attack.
By integrating with the industry-standard \work{Foundry} framework~\cite{foundry,DeFiHackLabs}, \sysname produces executable PoCs that can be readily used for downstream security analysis, such as root-cause investigation, and countermeasures deployment to mitigate financial losses.

Note that \sysname does not aim to detect attacks or identify root-cause vulnerabilities, instead, it focuses on synthesizing \poc from observed attack transactions. As such, our work is orthogonal to prior efforts on attack detection and root-cause analysis (see~\S~\ref{sec:relatedwork}). Moreover, unlike existing approaches that generate \poc by analyzing vulnerable contract code~\cite{foray,cpmmx,AdvSCanner,a1}, \sysname does not rely on access to contract source code, making it complementary to code-centric vulnerability analysis techniques.

\smalltitle{Threat Model and Scope.}
We consider an adversary whose primary motivation is financial gain.
Upon discovering a vulnerability, the adversary may initiate exploits through various vectors: direct interaction with victim contracts, deploying attack contracts, or executing complex multi-step transaction sequences.
We make no assumptions about the structure, transparency (\ie, open or closed source), or obfuscation level of the deployed attack contracts.

This work focuses on generating \poc code from attack transactions, including both launched transactions and pending transactions detected in the mempool.
Consistent with empirical observations that $\sim$95\% of DeFi exploits utilize an intermediary attack contract~\cite{lookahead,skyeye,BlockWatchdog,smartcat}, we assume by default that the transaction is initiated against an adversary-controlled contract (\S~\ref{sec:background:defi_attack}).
But \sysname is flexible: if an attack interacts directly with a victim contract, our synthesizer can be configured to generate direct calls to reproduce the exploit.
Currently, \sysname targets EVM-based blockchains (\eg, Ethereum, BNB Smart Chain), leveraging the maturity of EVM analysis tools~\cite{Panoramix,gigahorse,elipmoc,disco} and the high density of real-world incidents in these ecosystems~\cite{DeFiHackLabs}.

\subsection{Challenges and Solutions}

Synthesizing verifiable \poc from attack transactions poses challenges along two dimensions: recovering attack logic from low-level execution traces, and generating runnable, verifiable source code.

\textbf{(1) Extracting Attack Logic from Low-Level Traces.}
Execution traces (\eg, via \texttt{debug\_traceTransactions}~\cite{geth}) record every low-level operation during a transaction, but this granularity introduces substantial noise.
First, attack transactions involve complex interactions among adversary contracts, victim protocols, and peripheral components (\eg, tokens), making it non-trivial to isolate the attack-relevant sub-trace (\textbf{C1}).
Second, traces capture only a single concrete execution path: high-level control structures such as loops are fully unrolled, and concrete values lack symbolic semantics (\eg, \texttt{address(this)}).
These properties destroy the structural and semantic abstractions required for program recovery.
Existing EVM decompilers are designed for static bytecode analysis and cannot directly lift such dynamic, unrolled traces, while inherent challenges in precise memory modeling further hinder lifting (\textbf{C2}).

\textbf{(2) Synthesizing Verifiable PoC Code via LLMs.}
Although LLMs demonstrate strong general-purpose code generation capabilities~\cite{code_generation_survey,code_generation_survey2}, applying them to exploit reproduction introduces additional challenges.
First, exploit strategies vary significantly across DeFi protocols, requiring domain-specific knowledge and contextual reasoning that generic LLM prompts cannot easily capture (\textbf{C3}).
Second, LLMs are prone to hallucinations, resulting in syntactic errors or semantic deviations from the original exploit.
Ensuring that generated \poc code is not only plausible but also runnable and verifiable is therefore a central challenge (\textbf{C4}).

\begin{figure}
    \centering
    \includegraphics[width=0.99\linewidth]{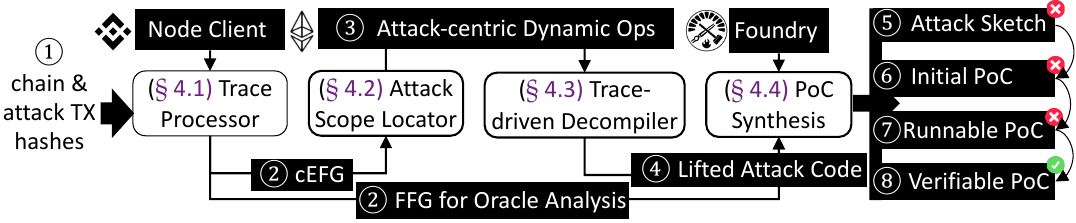}
    \caption{Workflow of \sysname.}
    \Description{}
    \label{fig:workflow}
\end{figure}

\smalltitle{\sysname Overview.}
To address these challenges, we design \sysname, a framework that automatically synthesizes verifiable \poc from attack transactions.
As shown in \autoref{fig:workflow}, \sysname first constructs a contract-centric execution representation to isolate attack-relevant behaviors (\circlednumber{1}-\circlednumber{3}), addressing trace noise and relevance (\textbf{C1}).
Then, it applies a trace-driven decompilation approach that fuses concrete execution evidence with symbolic lifting to recover high-level attack logic (\circlednumber{4}) despite unrolled control flow and dynamic semantics (\textbf{C2}).
To guide LLM-based synthesis, \sysname extracts attack-specific semantic context (\eg, contract names and fund flows) to draft structured \poc sketches (\circlednumber{5}), mitigating domain diversity (\textbf{C3}).
Finally, \sysname employs an exploit-aware refinement loop that prioritizes semantic exploitability over input–output equivalence (\textbf{C4}).
It first validates whether the generated \poc (\circlednumber{6}-\circlednumber{8}) reproduces unintended asset transfers, and then differentially refines execution behaviors to ensure verifiability and correctness.

\section{Detailed Design}\label{sec:methods}
This section presents the components of \sysname, detailing the transition from raw transaction traces to verifiable \poc code.

\subsection{Trace Processor}\label{sec:designs:tracer}
This component reconstructs full execution traces from raw transaction hashes and transforms low-level instruction streams into a structured representation suitable for semantic analysis.

\smalltitle{Contract-Centric Execution Flow Graph (cEFG).}
We introduce a \emph{contract-centric} Execution Flow Graph (cEFG), extending prior EFG designs~\cite{txspector} to capture cross-contract exploit logic compactly. 
Each node is a 5-tuple $\langle \textit{depth}, \textit{address}, \textit{call\_type}, \textit{input}, \textit{ins} \rangle$, representing the call context and sequential instructions, nodes sharing identical context are aggregated to reduce redundancy. 
Here, \textit{depth} is the call nesting level, \textit{address} identifies the executing contract, \textit{call\_type} indicates the invocation opcode (\eg, \ins{call}), \textit{input} captures calldata, and \textit{ins} lists instructions within that context.
To optimize the graph for exploit analysis, we exclude instructions in the node whose \textit{call\_type} is \texttt{STATICCALL}, as they do not modify contract states and are typically auxiliary to the core attack logic.
This structure preserves call-and-callback relationships and is particularly effective for multi-contract exploits.

\begin{figure}
\centering
\begin{tikzpicture}[
    x=0.28cm,
    y=0.5cm,
    codebox/.style={inner sep=0mm, text width=2.8cm, align=left},
    efg_node/.style={draw=black, thick, font=\sffamily\small, text=black, align=center},
    ins_style/.style={font=\bf\ttfamily\footnotesize, color=KeywordColor},
]

\node[efg_node] (n1) at (0, -1) {$e_1$};
\node[efg_node] (n2) at (0, -2) {$e_2$};
\node[efg_node] (n3) at (0, -3) {$e_3$};
\node[efg_node] (n4) at (0, -4) {$e_4$};
\node[efg_node] (n5) at (0, -5) {$e_5$};

\draw [-latex, thick, bend left=-45] (n1.west) to node[pos=0.1, left, ins_style] {STATICCALL} (n2.west);
\draw [-latex, thick, bend left=-45] (n2.west) to node[pos=0.1, left, ins_style] {DELEGATECALL} (n3.west);
\draw [-latex, thick, bend left=-45] (n3.west) to node[pos=0.1, left, ins_style] {RETURN} (n4.west);
\draw [-latex, thick, bend left=-45] (n4.west) to node[pos=0.1, left, ins_style] {RETURN} (n5.west);

\node[font=\sffamily\footnotesize, align=left, inner sep=1pt, right = 5mm of n2, yshift=-1mm] (e0) {depth: 1 \\ address: 0x3783c9... \\ call\_type: {\bf\ttfamily \textcolor{KeywordColor}{CALL}} \\ input: 0x0c96fa6200...\\ instructions: see right };

\node[codebox, right=3cm of n2, yshift=-2mm] (e1) {
\begin{lstlisting}[style=evm_style,linewidth=\linewidth, backgroundcolor={}]
0x0:PUSH1 0x80@\label{efg_e0_line1}@
0x2:PUSH1 0x40
0x4:MSTORE
0x5:PUSH1 0x4
0x7:CALLDATASIZE 0xc4@\label{efg_e0_line5}@
...
0x4f1:STATICCALL@\label{efg_e0_line12}@
\end{lstlisting}
};

\node[draw = black, dashed, inner sep = 0pt, fit=(e0)(e1)] (box1) {};

\draw[thick] (n1.north east) -- (box1.north west);
\draw[thick] (n1.south east) -- (box1.south west);

\node[font=\sffamily\footnotesize, align=left, inner sep=1pt, right = 5mm of n5, yshift=0mm] (e2) {instructions: see right };

\node[codebox, right=3cm of n5, yshift=0mm] (e3) {
\begin{lstlisting}[style=evm_style,linewidth=\linewidth, backgroundcolor={}]
0x4f2:ISZERO@\label{efg_e5_line1}@
...
\end{lstlisting}
};

\node[draw = black, dashed, inner sep = 0pt, fit=(e2)(e3)] (box2) {};

\draw[thick] (n5.north east) -- (box2.north west);
\draw[thick] (n5.south east) -- (box2.south west);
\end{tikzpicture}
\caption{The cEFG of WebkeyDAO attack.}
\label{fig:cefg}
\Description{}
\end{figure}
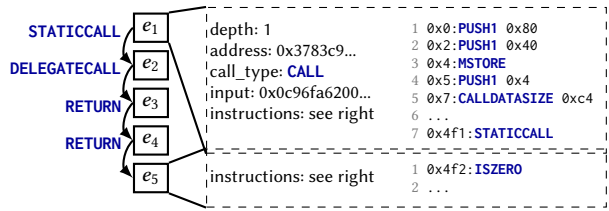

\textit{Example: cEFG of WebkeyDAO attack.} 
\autoref{fig:cefg} shows some nodes of a cEFG for a WebkeyDAO attack transaction. Execution starts at $e_1$, the root context, and creates a new node $e_2$ when a \texttt{STATICCALL} triggers a context switch. Returns from nested calls ($e_4$–$e_5$) to the previous context. $e_1$ and $e_5$ share the same context, and their instructions can be aggregated together for the \texttt{wheeaappP} invocation.

\smalltitle{Selective Argument Recording.}
Due to the dynamic semantics of the EVM, most notably data-dependent control dispatch (\eg, calldata-computed jump targets) and memory operations whose behaviors depend on runtime state (\eg, returndata produced by external calls), purely static analysis is insufficient to precisely reason about control-flow~\cite{skanf}, fund-flow~\cite{ObfProbe} and memory behavior (\eg, unoptimized memory in ~\autoref{fig:decompiled_attack_functions}).
Therefore, in addition to recording all executed instructions, we also record selected concrete runtime values to assist subsequent optimization and analysis.
Consequently, \sysname selectively records concrete values for 5 critical instruction categories required by the \textit{dual-decompiler} (\S~\ref{sec:designs:dual_decompiler}):
\begin{packeditemize} 
    \item \textbf{Constants:} Immediate values from \texttt{PUSH} operations (\eg, Line~\ref{efg_e0_line1}). 
    \item \textbf{Control Flow:} The conditional operands of \texttt{JUMPI} to facilitate control-flow graph reconstruction. 
    \item \textbf{External Context:} Instructions interacting with external contract states or code (\ie, \texttt{EXTCODECOPY}, \texttt{CODECOPY}, \texttt{CODESIZE}, and \texttt{EXTCODESIZE}), and the arguments of call-family instructions and their return values (\ie, \texttt{CALL}, \texttt{CALLCODE}, \texttt{DELEGATECALL}, \texttt{STATICCALL}, \texttt{RETURNDATASIZE}, and \texttt{RETURNDATACOPY}). 
    \item \textbf{Internal State:} Instructions accessing the contract’s persistent or transient storage, specifically \texttt{SLOAD} and \texttt{TLOAD}. 
    \item \textbf{Input Data:} Calldata-related instructions (\ie, \texttt{CALLDATACOPY}, \texttt{CALLDATASIZE}, and \texttt{CALLDATALOAD}) to track data flow from the transaction initiator. 
\end{packeditemize}

\smalltitle{Trace Implementation.} 
We implement a \textit{non-intrusive} customized tracer~\cite{custom-tracer} compatible with standard Ethereum clients (\eg, Geth~\cite{geth}). By hooking into the \texttt{step} function of the EVM, the tracer monitors execution in real-time. Upon detecting a context-switching opcode, the tracer dynamically instantiates a new cEFG node and begins capturing the execution metadata for the subsequent context. This approach enables high-fidelity dynamic analysis without requiring modifications to the underlying blockchain client source code.

\subsection{Attack Scope Localization}\label{sec:designs:locator}

To enable precise and scalable \poc synthesis, \sysname first localizes the \emph{attack scope} within a transaction trace.
Although a single attack transaction may involve many contracts, the exploit logic is typically confined to adversary-controlled contracts and a small number of their function invocations.
Concretely, we define the attack scope as the set of adversary-related contract functions and their executed instructions that collectively realize the exploit.

\begin{figure}
\centering
\begin{tikzpicture}[x=2cm,y=0.7cm,
    codebox/.style={fill=gray!5, inner sep=0mm, text width=2.8cm, align=left},
    attacknode/.style={draw=black, thick, inner sep=2pt, fill=red!10, align=center, font=\ttfamily\small},
    normalnode/.style={draw=black, thick, inner sep=2pt, fill=gray!10, align=center, font=\ttfamily\small},
    edgefont/.style={pos=0.5, draw=none, font=\bf\ttfamily\small, align=center, inner sep=1pt, color=KeywordColor},
    depth_style/.style={font=\sffamily\footnotesize, align=center, inner sep=1pt}
]
\node[attacknode] at (0,0) (n0) {A.wheeaappP};
\node[normalnode] at (-0.5, -1) (n1) {B.\_BASE\_TOKEN\_};
\node[normalnode] at (0.5, -1) (n2) {B.flashLoan};
\node[normalnode] at (-1, -2) (n3) {C.\_BASE\_TOKEN\_};
\node[normalnode] at (1, -2) (n4) {C.flashLoan};
\node[normalnode] at (-0.3, -3) (n5) {D.transfer};
\node[attacknode] at (1, -3) (n6) {A.DVMFlashLoanCall};
\node[normalnode, draw=none, fill=white] at (1.82, -3) (n7) {...};

\draw[-latex, left, thick] (n0) -- (n1) node[edgefont] {STATICCALL};
\draw[-latex, right, thick] (n0) -- (n2) node[edgefont] {CALL};
\draw[-latex, left, thick] (n1) -- (n3) node[edgefont] {DELEGATECALL};
\draw[-latex, left, thick] (n2) -- (n4) node[edgefont, pos=0.5] {DELEGATECALL};
\draw[-latex, left, thick] (n4) -- (n5) node[edgefont, pos=0.4] {CALL};
\draw[-latex, right, thick] (n4) -- (n6) node[edgefont, pos=0.5] {CALL};

\node[depth_style] at (-2, 0) (depth_1) {depth: 1};
\node[depth_style] at (-2, -1) (depth_2) {depth: 2};
\node[depth_style] at (-2, -2) (depth_3) {depth: 3};
\node[depth_style] at (-2, -3) (depth_4) {depth: 4};

\draw[dashed, thin] (depth_1.east) -- (n0);
\draw[dashed, thin] (depth_2.east) -- (n1);
\draw[dashed, thin] (depth_3.east) -- (n3);
\draw[dashed, thin] (depth_4.east) -- (n5);

\end{tikzpicture}
\caption{Call Graph. Adversary logic is in the red block.}
\label{fig:cg}
\Description{}
\end{figure}
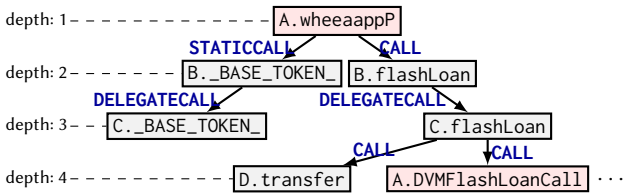

\smalltitle{Call Graph.}
\sysname constructs a hierarchical call graph ($CG$) by traversing the cEFG (\S~\ref{sec:designs:tracer}).
As shown in \autoref{fig:cg}, each node in the $CG$ represents a functional execution context, uniquely identified by the tuple $\langle \textit{address}, \textit{depth}, \textit{input} \rangle$.
This representation enables function-level reasoning even when contract source code or precise ABI information is unavailable.
Edges between nodes represent the \textit{call type}, preserving the inter-contract invocation hierarchy.

\smalltitle{Scope Identification Logic.}
Rather than relying on protocol-specific heuristics, \sysname localizes the attack scope based on execution semantics and adversarial control.
Specifically, exploit logic resides in execution contexts that are directly controlled by the attacker or derived from such contexts during execution.
Accordingly, \sysname performs a conservative traversal of the call graph, starting from the initial recipient contract $A_0$ of the attack transaction.
The attack scope is expanded based on the following criteria:
\begin{packeditemize}
    \item \textbf{Direct Invocations:} All functions of $A_0$ executed during the transaction.
    \item \textbf{Dynamic Instantiation:} Contracts dynamically deployed by $A_0$ or its descendants via \ins{CREATE} or \ins{CREATE2}.
    \item \textbf{Context Delegation:} Functions executed via \ins{DELEGATECALL} from any identified attack contract, as these executed external code is within the malicious contract's context.
\end{packeditemize}
This localization stage substantially reduces the analysis scope.
Compared to analyzing the full trace, localization decreases the number of executed instructions, inter-contract calls, and involved contracts by factors of 21$\times$, 9$\times$, and 13$\times$, respectively.
\sysname achieves a localization recall of \textbf{99.17\%}, with false negatives observed in only three attack incidents.
In these cases, \sysname generates exploitable PoCs by directly invoking the corresponding attack functions rather than reconstructing the attack contract code, which does not affect exploit reproduction correctness.

\smalltitle{Instruction Extraction.}
During call graph construction, each identified functional context is mapped back to its corresponding node in the cEFG.
Because each cEFG node is uniquely bound to an \textit{address}, \textit{input}, and \textit{depth}, \sysname incrementally aggregates all instruction blocks belonging to each attack-relevant function.
This process preserves internal state transitions and external call arguments, yielding complete instruction streams for subsequent trace-driven decompilation and \poc synthesis.

\textit{Example: CG of WebkeyDAO attack.} 
\autoref{fig:cg} shows the call graph built from the cEFG. The root node \texttt{A.wheeaappP} represents the initial invocation of the adversary contract \texttt{A} by the attacker. After this stage, we further locate the \texttt{DVMFlashLocanCall} function.

\subsection{Trace-driven Decompiler}\label{sec:designs:dual_decompiler}
\sysname addresses a fundamental gap between symbolic decompilation and execution tracing by introducing a \emph{trace-driven decompiler} that reconstructs high-level exploit semantics from a \textit{single} observed execution.
While symbolic decompilers recover structured code, they often blur concrete attack logic due to conservative abstractions, whereas execution traces precisely capture runtime behavior yet remain semantically opaque.
Building on this observation, \sysname fuses symbolic structure with trace-observed concrete values to deterministically lift an executed exploit into concise, attack-focused pseudocode.
At the core of this design are two tightly coupled components:
(i) a \emph{dual decompiler} (\S~\ref{sec:designs:decompilation}) that integrates concrete values into symbolic lifting, and
(ii) a \emph{trace compression} (\S~\ref{sec:designs:compressor}) that recovers iterative semantics from unrolled traces.

\subsubsection{Dual Decompiler}\label{sec:designs:decompilation}
We propose a \emph{dual decompiler}, a trace-driven semantic lifting framework that reconstructs high-level pseudocode from an already-executed trace with symbolic semantics.
Unlike static decompilation, which lacks concrete runtime information, and concolic or symbolic execution~\cite{cute,SYMBION}, which reason about multiple feasible paths via constraint solving, the dual decompiler deterministically lifts a single observed execution into a semantically meaningful representation.

The dual decompiler augments an existing symbolic decompiler with concrete runtime values recorded during tracing (\S~\ref{sec:designs:tracer}).
The design is engine-agnostic and requires no modification to the underlying lifting logic.
Conceptually, the dual decompiler operates through two synchronous processes: 
(i) \emph{symbolic lifting}, which recovers high-level structure using standard decompilation techniques, and 
(ii) \emph{concrete propagation}, which injects trace-observed values to refine and optimize the symbolic representation.

\smalltitle{Concrete Integration via Binding.}
The core of the dual decompiler is \emph{concrete integration}: symbolic expressions are explicitly anchored to concrete runtime values.
To realize this, we introduce a \emph{concrete-bind hook}. which extends the symbolic domain to a paired domain $\langle e, c \rangle$, where each symbolic expression $e$ is associated with its concrete value $c$ observed at the same execution point.
This binding enables concrete values to be propagated through the symbolic representation, while symbolic expressions retain semantic meaning for otherwise opaque constants.

Concrete integration immediately yields \textit{deterministic control-flow recovery}. 
Since branch conditions are bound to concrete values at each execution point, their concrete values are explicitly known during decompilation.
As a result, the decompiler follows the \textit{unique} branch, rather than speculating over alternative successors or exploring multiple paths. This trace-guided control-flow reconstruction ensures that the recovered control-flow graph precisely matches the observed exploit execution.

\smalltitle{Precise Memory and Local Variables Modeling.}
Concrete integration further enables precise and efficient modeling of memory and local variables.
Building on the memory abstraction introduced in \S~\ref{sec:background}, we re-model the memory into
\begin{equation*}
\setlength{\abovedisplayskip}{3pt}
\setlength{\belowdisplayskip}{3pt} 
\begin{array}{l}
D_M^c : (\mathit{range^c}, \mathit{len^c}) \mapsto \langle e, c \rangle,
\end{array}
\end{equation*}
where the memory region $(\mathit{range^c}, \mathit{len^c})$ is represented using concrete values.
This design choice simplifies memory access semantics and enables deterministic resolution of \textit{reads} and \textit{writes} during decompilation. For memory writes, the accessed memory range $(r, \ell)$ is concretely determined by the execution.
Concrete ranges may partially overlap; however, using deterministic segmentation of overlapping concrete segments
(as implemented in existing static analysis tools, \eg, the \texttt{split\_*} functions in \work{Panoramix}~\cite{Panoramix}), all memory accesses can be mapped to non-overlapping segments for reads and writes. 
By operating on these concrete segments, the dual decompiler deterministically resolves memory accesses
without introducing aliasing ambiguities, and without requiring symbolic pointer reasoning or constraint solving.

Local variables are modeled analogously using a variable map
\begin{equation*}
\setlength{\abovedisplayskip}{3pt}
\setlength{\belowdisplayskip}{3pt} 
\begin{array}{l}
D_V^c : \mathit{var} \mapsto \langle e, c \rangle,
\end{array}
\end{equation*}
which is updated and queried deterministically.

\begin{algorithm}[t]
    \centering
    \caption{Trace-driven Dual Decompiler}
    \label{alg:semantic_interpretation}
    \small
    \begin{algorithmic}[1]
    \Statex\underline{$\textsc{DualDecompiler}(P, D_V^c, D_M^c)$}
    \State $P' \gets [~]$
    \Comment{\footnotesize output: high-level pseudocode with concrete bindings}

    \For{\textit{each} statement $s \in P$}

        \If{$s.\textsf{type} = \textsf{setmem}$}
            \State $(r, \ell) \gets \textsc{ConcreteRange}(s)$
            \Comment{\footnotesize concrete memory range }
            \State $(e, c) \gets \textsc{Lift}(s.\textsf{value}, D_V^c, D_M^c)$
            \Comment{\footnotesize lift value with concrete-bind}
            \State $D_M^c[(r, \ell)] \gets \langle e, c \rangle$
            \Comment{\footnotesize update memory map deterministically}

        \ElsIf{$s.\textsf{type} = \textsf{setvar}$}
            \State $(e, c) \gets \textsc{Lift}(s.\textsf{value}, D_V^c, D_M^c)$
            \State $D_V^c[s.\textsf{var}] \gets \langle e, c \rangle$
            \Comment{\footnotesize update local variable map}

        \ElsIf{$s.\textsf{type} = \textsf{if}$}
            \State $(e_{cond}, \_) \gets \textsc{Lift}(s.\textsf{cond}, D_V^c, D_M^c)$ \Comment{\footnotesize deterministic branch}
            \State $B \gets \textsc{DualDecompiler}(s.\textsf{branch}, D_V^c, D_M^c)$
            \State $P'.\textsf{append}(\langle \textsf{if}, e_{cond}, B \rangle)$
            \Comment{\footnotesize use symbolic condition semantic $e_{cond}$}
        \Else \Comment{\footnotesize handle other statement types: calls, arithmetic, etc.}
            \State $(e, \_) \gets \textsc{Lift}(s, D_V^c, D_M^c)$
            \State $P'.\textsf{append}(e)$
        \EndIf
    \EndFor
    \State \Return $P'$
    \Comment{\footnotesize final pseudocode}
    \end{algorithmic}
\end{algorithm}

\smalltitle{Pseudocode Generation.}
With concrete control-flow and precise memory and local-state modeling in place, the dual decompiler performs trace-driven semantic lifting to reconstruct high-level pseudocode.
Algorithm~\ref{alg:semantic_interpretation} summarizes this process.
The algorithm sequentially interprets trace statements, lifts values via the concrete-bind hook,
updates $D_M^c$ and $D_V^c$ deterministically, and emits symbolic expressions annotated with concrete bindings.

Overall, the dual decompiler reconstructs semantically meaningful pseudocode from a single, concrete execution trace under symbolic structure.
It neither generalizes beyond observed behavior nor speculates on unexecuted paths, distinguishing it fundamentally from symbolic or concolic execution.
\autoref{fig:dec_attack_functions_in_trace} shows the optimized decompiled pseudocode.
Compared with existing outputs (see~\autoref{fig:decompiled_attack_functions}), temporary variables and memory references (\texttt{Mem}) are eliminated, significantly enhancing code readability.

\subsubsection{Trace Compression}\label{sec:designs:compressor}

Execution traces fully unroll loops, producing linear sequences that can obscure high-level logic and inflate token consumption for downstream LLM-based synthesis (evaluated in~\S~\ref{sec:eval:ablation_studies}).
To mitigate this, we introduce a lightweight \emph{trace compression} module to summarizes repetitive structures.
Unlike traditional static analysis that recovers loops from a CFG~\cite{LoopTerminology}, this module leverages the \textit{deterministic iteration counts} inherent in the trace to accurately identify and parameterize loops.

\smalltitle{Deterministic Loop Identification.}
Let a trace be represented as a sequence of concrete program counters $\mathcal{T} = [\textsf{PC}_1, \dots, \textsf{PC}_n]$.
Loop identification aims to locate the \textit{loop entry} $e_\ell$, \textit{exit} $x_\ell$, the exact number of \textit{iterations} $k_\ell$, and the corresponding \textit{loop bodies} $\mathcal{B}_\ell = [B_1, \dots, B_{k_\ell}]$.
We detect consecutively repeated PC subsequences within $\mathcal{T}$, which deterministically define $e_\ell$, $x_\ell$, and $k_\ell$,
providing concrete loop bodies $B_i$ for subsequent summarization.
Formally, a repeated subsequence $S = [\textsf{PC}_p, \dots, \textsf{PC}_q]$ is recognized as a loop body if $\exists m \ge 2$ such that $S$ occurs consecutively $m$ times in $\mathcal{T}$.

\smalltitle{Parameterized Loop Summarization.}
Once loop bodies $\mathcal{B}_\ell$ are identified, repeated statements are abstracted into a single \emph{parameterized template} $\hat{B}_\ell$.
We perform a pairwise diff between adjacent loop bodies to classify operations as either invariant or variant. 
Specifically, for each statement $s$ in the loop body:

\begin{packeditemize}
    \item \textbf{Invariant:} If $s$ is identical across all iterations, it is retained directly in the loop template $\hat{B}_\ell$ and executed once per iteration, controlled by a loop index variable $v_\ell$ (\eg, \texttt{i}).
    \item \textbf{Deterministic Variation:} If $s$ varies according to a predictable function of the loop index (\eg, sequential memory addresses, array indices, or arithmetic over $v_\ell$), we represent $s$ as a deterministic function $f_s(v_\ell)$ within the loop template.
    \item \textbf{Complex Divergence:} If $s$ exhibits irregular or non-monotonic changes, we introduce a temporary symbolic variable $t_s$ (\eg, \texttt{arr1}) and express the statement in the template as $s[t_s, v_\ell]$ (\eg, \texttt{arr1[i]}), preserving the data flow across iterations.
\end{packeditemize}
Formally, the parameterized loop template is
\begin{equation*}
\setlength{\abovedisplayskip}{3pt}
\setlength{\belowdisplayskip}{3pt} 
\begin{array}{l}
\hat{B}_\ell = \{ s \mapsto \textsf{Invariant} \cup f_s(v_\ell) \cup s[t_s, v_\ell] \mid s \in B_1 \},
\end{array}
\end{equation*}
where each $s \in B_1$ corresponds to the first statement in $\mathcal{B}_\ell$.

This trace-driven abstraction produces a compact representation that preserves the semantic behavior of the original execution.

\subsection{PoC Synthesiser}\label{sec:designs:synthesiser}
The final stage of \sysname leverages LLMs to transform the optimized pseudocode into a verifiable source-level \poc. To ensure reliability, we decouple the synthesis process into \textit{Sketch Generation} (\S~\ref{sec:designs:sketch}) and \textit{Iterative Refinement} (\S~\ref{sec:designs:refinement}).

\begin{figure}
\centering
\begin{minipage}[b]{\linewidth}
\begin{lstlisting}[style=solidity_style,linewidth=\columnwidth]
pragma solidity ^0.8.10; /*!\label{poc_line:1}!*/
import "forge-std/Test.sol"; // foundry test framework /*!\label{poc_line:2}!*/
import "../interface.sol"; // import some implemented interfaces /*!\label{poc_line:3}!*/
/*!\rulebox{\textbf{$\langle$Constant Addresses$\rangle$}}!*/ // e.g., address constant attacker=0x...;/*!\label{poc_line:4}!*/
contract ContractTest is Test {
  function setUp() public {  /*!\label{poc_line:6}!*/
    vm.createSelectFork("/*!\rulebox{\textbf{$\langle$chain$\rangle$}}!*/", /*!\rulebox{\textbf{$\langle$blocknumber$\rangle$}}!*/-1); /*!\label{poc_line:7}!*/ // fork chain states
    /*!\rulebox{\textbf{$\langle$Balances Assumptions$\rangle$}}!*/ // e.g., deal(attacker, 1 ether); /*!\label{poc_line:8}!*/
  }    /*!\label{poc_line:9}!*/
  function testPoC() public {
    /*!\insbox{\textbf{$\langle$Pre-Attack Balances$\rangle$}}!*/ /*!\label{poc_line:12}!*/ //e.g., emit log_named_uint("before attack", ...);
    vm.startPrank(attacker); // set the msg.sender as attacker
    AttackerC attC = new AttackerC(); // create attacker contract
    /*!\rulebox{\textbf{$\langle$Attack Calls$\rangle$}}!*/ /*!\label{poc_line:14}!*/ // e.g., attC.attack{value: ...}(...);
    vm.stopPrank();
    /*!\insbox{\textbf{$\langle$Post-Attack Balances$\rangle$}}!*/ /*!\label{poc_line:17}!*/ //e.g., emit log_named_uint("after attack", ...);
  }
}
contract AttackerC { // the core attack logic implementation
  /*!\rulebox{\textbf{$\langle$Attack Functions$\rangle$}}!*/ { /*!\llmbox{\textbf{$\langle$Attack Logic$\rangle$}}!*/ } // e.g., function attack(...)
  /*!\llmbox{\textbf{$\langle$Other Functions$\rangle$}}!*/ // e.g., callback functions /*!\label{poc_line:23}!*/
}
/*!\llmbox{\textbf{$\langle$Other Contracts$\rangle$}}!*/ // e.g., helper contracts
/*!\rulebox{\textbf{$\langle$Involved ABIs$\rangle$}}!*//*!\label{poc_line:24}!*/
\end{lstlisting}
\begin{tikzpicture}[remember picture,overlay]
\node[draw, align=left, font=\scriptsize, dashed] at (7,3.4em) {\rulebox{$\langle$...$\rangle$} Provided Attack Context \\ \insbox{$\langle$...$\rangle$} Instrumented Oracle Logs\\ \llmbox{$\langle$...$\rangle$} PLACEHOLDER for LLMs};
\end{tikzpicture}
\vspace{-1em}
\caption{\poc Sketch Template in \work{DeFiHackLabs}~\cite{DeFiHackLabs} Style.}
\label{fig:poc-template}
\end{minipage}
\Description{PoC Template Provided to LLMs.}
\end{figure}

\subsubsection{Sketch Synthesis} 
\label{sec:designs:sketch}

Directly prompting an LLM to generate exploit code is often infeasible due to: \textit{(1) Safety Alignment}: LLMs' built-in mechanisms may refuse to produce malicious payloads\footnote{Although some jailbreak strategies~\cite{masterkey} were proposed, this is not our focus.}; and \textit{(2) Framework Dependencies}: \poc in modern testing framework (\eg, \work{Foundry}~\cite{foundry}) require precise environmental configuration and metadata (\eg, block forks), which are difficult for LLMs to infer implicitly.
Rather than treating the LLM as a standalone code generator, \sysname adopts a \emph{sketch-based synthesis} strategy that decouples \emph{analysis} from \emph{code completion}.

\smalltitle{Sketch Components.}
At a high level, \sysname first derives a structured \poc sketch (see~\autoref{fig:poc-template}) from dynamic execution evidence and then constrains the LLM to complete only well-defined placeholders, which both reduces reliance on unconstrained LLM generation and ensures compatibility with real-world exploit testing environments.
The sketch comprises three main components: 
\begin{packeditemize}
    \item \textbf{Rule-derived Attack Context:} Pre-configured execution environment data tailored to the specific attack (\eg, Line~\ref{poc_line:7}).
    \item \textbf{Instrumented Oracle Logs:} Logic-based logs automatically embedded into the harness for exploit verification (\ie, Lines~\ref{poc_line:12}, ~\ref{poc_line:17}).
    \item \textbf{Functional Placeholders:} Structural stubs that the LLM populates using the decompiled pseudocode (\eg, Line~\ref{poc_line:23}).
\end{packeditemize}
Specifically, the Rule-derived Attack Context encompasses:  \textit{(i) Environmental Configuration} (Lines~\ref{poc_line:6}-\ref{poc_line:9}), which defines the blockchain network (\eg, Ethereum, BSC), the specific fork block number, and the minimum funding requirements derived from fund-flow analysis; 
\textit{(ii) Metadata and Interfaces}, including target contract addresses, human-readable aliases (Line~\ref{poc_line:4}), and essential ABI specifications (Line~\ref{poc_line:24}); 
and \textit{(iii) Invocation Sequence} (Line~\ref{poc_line:14}), which provides a high-level roadmap of the function calls the \poc must execute.

\smalltitle{Fund Flow and Oracle Extraction.}
Correctness of a synthesized \poc is defined by whether it reproduces the \emph{semantic outcome} of the original exploit, rather than merely executing without errors.
In DeFi attacks, such outcomes manifest as observable \emph{economic revenue} (\ie, asset redistribution) across accounts, which can be precisely recovered from execution traces.
\sysname models the fund flow of a transaction as net balance changes over all involved assets.
Formally, for an account $a$ and an asset $t$ (including ether and ERC20 tokens), we define
\begin{equation*}
\setlength{\abovedisplayskip}{3pt}
\setlength{\belowdisplayskip}{3pt} 
\begin{array}{l}
\Delta(a, t) = \textit{balance}_{\textit{post}}(a, t) - \textit{balance}_{\textit{pre}}(a, t).
\end{array}
\end{equation*}
Non-zero $\Delta(a, t)$ values collectively characterize the economic effect of the transaction. \sysname first evaluates the net profit of the sender (\ie, attacker) and the attack contract. If no significant gain is detected, we iterate through all involved addresses to identify the entity with the maximum net profit as the presumed beneficiary.

To extract fund flow, \sysname monitors trace instructions that induce asset transfers.
Token movements are identified via \texttt{LOG}-based events (\ie, \texttt{Transfer}, \texttt{Deposit}, \texttt{Withdrawal}),
while native ether transfers are captured through value-carrying instructions (\eg, \texttt{CALL}, \texttt{CREATE(2)}, and \texttt{SELFDESTRUCT}).
The extracted fund flow is used to derive the minimum attacker balance (Line~\ref{poc_line:8}) and to synthesize balance-change assertions that serve as deterministic oracles for exploit validation (\ie, Lines~\ref{poc_line:12}, \ref{poc_line:17}).

\smalltitle{LLM-driven Completion.}
Finally, \sysname provides the LLM with the ordered, decompiled pseudocode (\S~\ref{sec:designs:dual_decompiler}) and instructs it to fill the functional placeholders. To ensure semantic fidelity during \ins{DELEGATECALL} operations, where logic is executed within the caller's state context, we explicitly prompt the LLM to encapsulate such logic within distinct contract structures. This ensures the generated code faithfully preserves the runtime behavior of the original exploit.

\subsubsection{Exploit-Aware Refinement} \label{sec:designs:refinement}

LLM-generated \poc code may suffer from both syntactic errors and semantic deviations.
To ensure verifiability, \sysname introduces an \emph{exploit-aware iterative refinement loop} that differs fundamentally from prior program repair or differential testing approaches.
Instead of enforcing strict input–output equivalence, our refinement prioritizes \emph{semantic exploitability}, \ie, whether the synthesized \poc reproduces the unintended asset transfer observed in the original attack.

\smalltitle{Syntax Refinement.}
We first eliminate syntactic inconsistencies using compiler-directed feedback.
Upon compilation failure, \sysname extracts error messages from the \work{Foundry} environment and constructs a focused repair prompt containing only PoC snapshot and the reported errors.
Each refinement iteration is stateless, avoiding prompt accumulation and reducing token overhead.
This process repeats until the \poc successfully compiles.

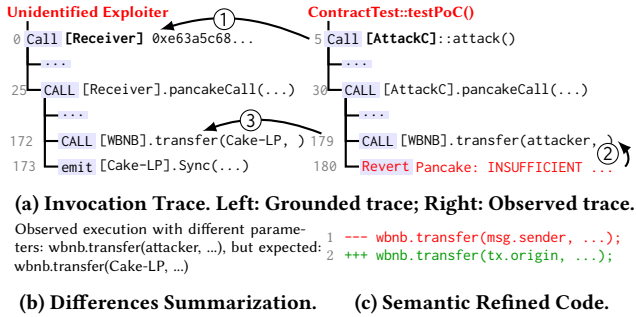
\begin{figure}[t]
\centering
\begin{subfigure}[b]{\linewidth}
\begin{tikzpicture}[x=0.23cm,y=0.34cm,
    attacker/.style={anchor=west, font=\sffamily\scriptsize, align=center, inner sep=0pt},
    calltype/.style={anchor=west, font=\sffamily\scriptsize, align=center, inner sep=1pt, fill=blue!10},
    nodetext/.style={anchor=west, font=\ttfamily\scriptsize, align=center, inner sep=1pt},
    toggle/.style={anchor=west, draw=black, fill=white, inner sep=0pt, font=\tiny, minimum size=3pt},
    coord/.style={font=\scriptsize, align=center, inner sep=0pt},
    lineno/.style={anchor=east, font=\ttfamily\scriptsize, align=center, inner sep=1pt, text=gray},
    funcbox/.style={draw=red, thick, rounded corners, inner sep=2pt, fill=yellow!10},
    codebox/.style={fill=gray!5, inner sep=0mm, text width=6.5cm, align=left},
    line_style/.style={-latex, thick},
    numberstyle/.style={anchor=east, font=\ttfamily\small, align=center, inner sep=0pt}
]
  \begin{scope}[xshift=0cm]
    \node[attacker] at (0, 0) (a1_attacker) {\textcolor{red}{\textbf{Unidentified Exploiter}}};
    \node[calltype] at (1, -1) (a1_1call) {$\texttt{Call}$};
    \node[nodetext, right=0mm of a1_1call.east] (a1_2callt) {\textcolor{black}{\textbf{[Receiver]} 0xe63a5c68...}};
    \node[calltype] at (2, -2) (a1_2call) {$\texttt{...}$};
    \node[calltype] at (2, -3) (a1_3call) {$\texttt{CALL}$};
    \node[nodetext, right=0mm of a1_3call.east] {\textcolor{black}{[Receiver].pancakeCall(...)}};
    \node[calltype] at (3, -4) (a1_4call) {$\texttt{...}$};
    \node[calltype] at (3, -5) (a1_5call) {$\texttt{CALL}$};
    \node[nodetext, right=0mm of a1_5call.east] (a1_5callt) {\textcolor{black}{[WBNB].transfer(Cake-LP, )}};
    \node[calltype] at (3, -6) (a1_6call) {$\texttt{emit}$};
    \node[nodetext, right=0mm of a1_6call.east] {\textcolor{black}{[Cake-LP].Sync(...)}};

    \foreach \k/\i in {1/0, 2/{}, 3/25, 4/{}, 5/172, 6/173} {
      \node[lineno] at ($(0,0)+(2mm+\k*0.4mm,-\k*3.4mm)$) {\i};
    }

    \draw[-, thick] ([xshift=-2mm]a1_1call.north) -- ($([xshift=-2mm]a1_1call.north |- a1_attacker.south)$);
    \draw[-, thick] ([xshift=-2mm]a1_1call.south) |- (a1_2call);
    \draw[-, thick] ([xshift=-2mm]a1_1call.south) |- (a1_3call);
    \draw[-, thick] ([xshift=-2mm]a1_3call.south) |- (a1_4call);
    \draw[-, thick] ([xshift=-2mm]a1_3call.south) |- (a1_5call);
    \draw[-, thick] ([xshift=-2mm]a1_3call.south) |- (a1_6call);
  \end{scope}

  \begin{scope}[xshift=40mm]
    \node[attacker] at (0, 0) (a2_attacker) {\textcolor{red}{\textbf{ContractTest::testPoC()}}};
    \node[calltype] at (1, -1) (a2_1call) {$\texttt{Call}$};
    \node[nodetext, right=0mm of a2_1call.east] (a2_2callt) {\textcolor{black}{\textbf{[AttackC]}::attack()}};
    \node[calltype] at (2, -2) (a2_2call) {$\texttt{...}$};
    \node[calltype] at (2, -3) (a2_3call) {$\texttt{CALL}$};
    \node[nodetext, right=0mm of a2_3call.east] {\textcolor{black}{[AttackC].pancakeCall(...)}};
    \node[calltype] at (3, -4) (a2_4call) {$\texttt{...}$};
    \node[calltype] at (3, -5) (a2_5call) {$\texttt{CALL}$};
    \node[nodetext, right=0mm of a2_5call.east] (a2_5callt) {\textcolor{black}{[WBNB].transfer(attacker, )}};
    \node[calltype] at (3, -6) (a2_6call) {\textcolor{red}{$\texttt{Revert}$}};
    \node[nodetext, right=0mm of a2_6call.east] (a2_6callt) {\textcolor{red}{Pancake: INSUFFICIENT ...}};

    \foreach \k/\i in {1/5, 2/{}, 3/30, 4/{}, 5/179, 6/180} {
      \node[lineno] at ($(0,0)+(2mm+\k*0.4mm,-\k*3.4mm)$) {\i};
    }

    \draw[-, thick] ([xshift=-2mm]a2_1call.north) -- ($([xshift=-2mm]a2_1call.north |- a2_attacker.south)$);
    \draw[-, thick] ([xshift=-2mm]a2_1call.south) |- (a2_2call);
    \draw[-, thick] ([xshift=-2mm]a2_1call.south) |- (a2_3call);
    \draw[-, thick] ([xshift=-2mm]a2_3call.south) |- (a2_4call);
    \draw[-, thick] ([xshift=-2mm]a2_3call.south) |- (a2_5call);
    \draw[-, thick] ([xshift=-2mm]a2_3call.south) |- (a2_6call);
  \end{scope}

  \draw[line_style, bend left=-20] ([xshift=-2mm]a2_1call.west) to node[midway, above, numberstyle] {\circlednumber{1}} (a1_2callt.north);
  \draw[line_style] ([xshift=1mm]a2_6callt.east) to[out=45, in=-45] node[midway, xshift=-1mm, numberstyle] {\circlednumber{2}} (a2_5callt.east);
  \draw[line_style, bend left=-20] ([xshift=-5mm,yshift=1mm]a2_5call.west) to node[midway, numberstyle] {\circlednumber{3}} (a1_5callt.north);

\end{tikzpicture}

\caption{Invocation Trace. Left: Grounded trace; Right: Observed trace.}
\label{fig:observed_trace_of_Unidentified_exploit}
\end{subfigure}
\\
\begin{subfigure}[b]{0.23\textwidth}
\begin{minipage}[b]{\linewidth}
\begin{tikzpicture}
    \node[font=\sffamily\scriptsize, align=left, inner sep=0pt, text width=5cm] at (0,0) {Observed execution with different parameters: wbnb.transfer(attacker, ...), but expected: wbnb.transfer(Cake-LP, ...) };
\end{tikzpicture}
\end{minipage}
\caption{Differences Summarization.}
\label{fig:differences_summarization}
\end{subfigure}
\begin{subfigure}[b]{0.23\textwidth}
\begin{minipage}[b]{\linewidth}
\begin{lstlisting}[style=solidity_style,linewidth=\columnwidth, breaklines=true]
--- wbnb.transfer(msg.sender, ...);
+++ wbnb.transfer(tx.origin, ...);
\end{lstlisting}
\end{minipage}
\caption{Semantic Refined Code.}
\label{fig:semantic_refined_code}
\end{subfigure}
\caption{Illustration for Error-Directed Trace Alignment.}
\label{fig:illustration_of_semantic_refinement}
\Description{Socket Incident Example.}
\end{figure}

\smalltitle{Semantic Refinement.}
A compilable \poc does not necessarily imply exploitability.
\sysname therefore performs semantic refinement in a \emph{two-tier, exploit-driven manner}:
(1) \textit{Profit-Oriented Validation.}
We first evaluate whether executing the synthesized \poc yields the unintended asset transfer as the ground-truth attack. 
This profit oracle serves as a coarse-grained semantic criterion: if satisfied, the \poc is considered semantically correct, even if low-level execution details differ.
(2) \textit{Execution Alignment.}
If the profit oracle fails, \sysname performs a targeted comparison between the execution trace produced by \work{Foundry} and the original on-chain trace.
Rather than enforcing full trace equivalence, we localize discrepancies that are likely to affect exploitability, such as mismatched call targets, incorrect addresses, or inconsistent argument values, and translate them into actionable feedback for the LLM.

\smalltitle{Greedy, Error-Directed Trace Alignment.} 
Direct trace comparison is challenging due to benign divergences caused by contract creation, address randomization, or call reordering.
To robustly localize semantic mismatches, as illustrated in \autoref{fig:illustration_of_semantic_refinement}, \sysname adopts a \textit{greedy, error-directed} alignment strategy consisting of three steps:
\circlednumber{1} we align contract addresses in the two traces according to their creation order to establish a coarse correspondence (\eg, \texttt{AttackC} and \texttt{Receiver}). \circlednumber{2} we greedily prioritize locations where Foundry reports runtime errors (\eg, Line~180) and attempt repairs beginning at those error sites. \circlednumber{3} To infer the semantics that should have been executed at a problematic location, we perform a local, windowed matching: within a small neighborhood of the error, we compare call targets, argument patterns, and other call-site metadata to identify the call in the generated trace that most closely matches the expected operation. We then present the paired statements (the expected and the observed calls) to the LLM as focused context for semantic correction (\eg, \autoref{fig:differences_summarization}). 
This combination of creation-order alignment, error-directed greedy repair, and localized window matching helps localize semantic differences and guides effective LLM-driven corrections (\eg, \autoref{fig:semantic_refined_code}).

By combining syntax and semantic refinement, \sysname guarantees that the automatically generated \poc code is both compilable and semantically faithful, accurately reproducing the attack logic while remaining compatible with Foundry.

\section{Evaluation}\label{sec:evaluation}

In this section, we evaluate \sysname to demonstrate its effectiveness and efficiency in synthesizing executable \poc from low-level execution traces.
As \textit{no prior work} directly addresses this task, we compare \sysname with representative alternative approaches for exploit analysis and \poc generation, and conduct ablation studies to assess the contribution of each design component.
Specifically, our experiments focus on the following research questions:
\begin{packeditemize}
\item \textbf{(RQ1)} How effective and efficient is \sysname in generating \poc for real-world attacks (\S~\ref{sec:eval:main_result}-\S~\ref{sec:eval:failure_analysis})?
\item \textbf{(RQ2)} To what extent does each design component contribute to the overall effectiveness (\S~\ref{sec:eval:ablation_studies})?
\item \textbf{(RQ3)} How does \sysname compare against alternative methods for exploit reproduction and \poc generation (\S~\ref{sec:eval:comparison})?
\end{packeditemize}

\subsection{Datasets and Experimental Setup}

\smalltitle{Datasets.} 
To evaluate \sysname, we constructed \work{TracEXP-ds}, a large-scale benchmark comprising \textbf{321} real-world attack incidents and their corresponding transaction hashes from the ETH and BSC networks.
The dataset spans 20 months (January 2024 to July 2025).
Compared to existing benchmarks, such as \work{Foray}~\cite{foray} (34 incidents) and \work{A1}~\cite{a1} (38 incidents), \work{TracEXP-ds} is an order of magnitude larger.
This scale is achieved because \sysname is agnostic to attack types, and the low computational cost of our pipeline enables high-throughput evaluation. The dataset is curated from two sources:

\begin{packeditemize}
\item \textbf{Exploits with Ground-Truth PoCs (\textit{w-poc}):} We collected 178 incidents from \work{DeFiHackLabs}~\cite{DeFiHackLabs}. We extracted ``attack transaction hashes'' primarily from analyst metadata, and used exploit reports as a fallback when unavailable. After filtering two cases (\href{https://github.com/SunWeb3Sec/DeFiHackLabs/blob/main/src/test/2024-02/EGGX_exp.sol}{EGGX} and \href{https://github.com/SunWeb3Sec/DeFiHackLabs/blob/main/src/test/2024-01/LQDX_alert_exp.sol}{LQDX}) due to missing transaction records, we retained \textbf{176} incidents as our ground-truth set. 
\item \textbf{Exploits without Existing PoCs (\textit{wo-poc}):} To evaluate the generality of \sysname on in-the-wild attacks, we aggregated over 300 alerts from prominent security firms, including CertiK~\cite{CertiKAlert}, SlowMist~\cite{SlowMist_Team}, and TenArmor~\cite{TenArmorAlert}. After deduplication against the \textit{w-poc} set, we retained \textbf{145} unique incidents.
\end{packeditemize}

\smalltitle{Implementation.}
\sysname utilizes \work{Panoramix}~\cite{Panoramix} as the core decompiler backend, adopting its output as the high-level Intermediate Representation (IR) for subsequent synthesis.
Nevertheless, the modular design of \sysname ensures that it remains decompiler-agnostic and can be readily adapted to other decompilers~\cite{gigahorse,elipmoc,disco}.
We employ \work{GPT-5}~\cite{gpt5} as the primary LLM for \poc synthesis,
given its strong performance on recent code reasoning and generation tasks~\cite{hu2026linecontextrepositorylevelcode,lang2026perishflourishholisticevaluation,smartpoc}.
We accessed the model via the \work{OpenRouter}~\cite{openrouter} unified API. In accordance with established best practices for software engineering tasks~\cite{using_gpt5}, we optimized the model parameters by setting \texttt{reasoning\_effort} to \textit{minimal} and \texttt{verbosity} to \textit{low}.

\smalltitle{Experimental Setup.}
To ensure a consistent and bounded evaluation, we imposed a 10-minute timeout for the trace processing and \poc sketch generation phases.
For the iterative refinement loop, we set a maximum budget of 5 iterations for syntactic repair and 3 iterations for semantic refinement.
During the semantic refinement stage, if a correction introduces new syntax errors, we permit up to 3 auxiliary syntactic repair attempts per semantic iteration.
All experiments were conducted on an Ubuntu 22.04 server equipped with an Intel Xeon Gold 6252 CPU (2.10 GHz) and 251 GB of RAM.

\subsection{Main Results for \sysname} \label{sec:eval:main_result}

To assess the effectiveness of \sysname, we evaluate the generated \poc along three dimensions:

\begin{packeditemize}
    \item \textit{Readable}: \sysname successfully produces human-readable \poc source code from attack transactions within a fixed time budget;
    \item \textit{Runnable}: the generated source code compiles successfully under the \work{Foundry} framework;
    \item \textit{Verifiable}: the compiled \poc not only runs but also reproduces the original attack by satisfying the predefined profit oracles.
\end{packeditemize}
While \textit{verifiability} is our primary objective, \textit{readable} and \textit{runnable} outputs are also practically valuable, as they provide structured templates that substantially reduce the manual effort required for exploit analysis and reproduction (see \S~\ref{sec:discussion}).

\begin{figure}
    \centering
    \begin{subfigure}{0.49\linewidth}
        \centering
        \includegraphics[width=\textwidth]{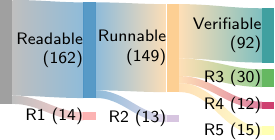}
        \caption{For 176 incidents with \poc.}
        \label{fig:eval:main_result:w_pocs}
    \end{subfigure}
    \hfill
    \begin{subfigure}{0.49\linewidth}
        \centering
        \includegraphics[width=\textwidth]{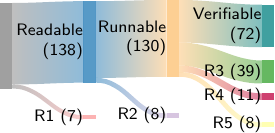}
        \caption{For 145 incidents without \poc.}
        \label{fig:eval:main_result:wo_pocs}
    \end{subfigure}
    \caption{Results of \sysname on \dataset. R1-R5 donote the reasons for unsuccessful cases: \emph{Timeout}, \emph{Syntax Repair Failure}, \emph{Incomplete Context}, \emph{Execution Failure}, and \emph{Semantic Repair Failure}, respectively.}
    \label{fig:eval:main_result}
    \Description{}
\end{figure}

\smalltitle{Overall Results.}
\autoref{fig:eval:main_result} summarizes the evaluation results on \dataset.
Out of 321 exploit incidents, \sysname produces 300 (93.46\%) \textit{readable} \poc codes.
Among them, 279 (93.00\%) are \textit{runnable}, \ie, they compile successfully under the \work{Foundry} framework without manual intervention.
Finally, 164 of the runnable \poc codes (58.78\%) are \textit{verifiable}, successfully reproducing the original attack by satisfying the predefined profit oracles.

\smalltitle{Real-World Impact.}
To assess the practical utility of \sysname, we anonymously submitted 33 generated \poc codes from the \textit{wo-poc} subset to the \work{DeFiHackLabs}~\cite{DeFiHackLabs} 2025 Summer Contest\footnote{The final leaderboard is available at \href{https://leaderboardhq.com/5kmsevrj}{https://leaderboardhq.com/5kmsevrj}.}.
These submissions accounted for 38\% of all community contributions in August 2025.
Notably, the number of \poc submissions generated by \sysname exceeded the 31-day output of five top human contributors by factors ranging from 1.32$\times$ to 33.00$\times$. Measured on a per-attack-event basis, \sysname generated \poc submissions 16.50$\times$ to 33.00$\times$ more efficiently than the other 4 contributors.
This result provides external evidence that \sysname can efficiently generate practically usable \poc artifacts at scale, both earning positive feedback from the community and also a \textbf{\$900 bounty}.

\begin{figure}[t]
    \centering
    \begin{subfigure}[b]{0.49\linewidth}
        \centering
        \includegraphics[width=\linewidth]{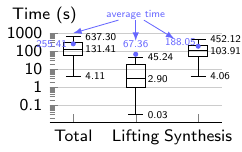}
        \caption{Time Cost.}
        \label{fig:eval:time_cost}
    \end{subfigure}
    \hfill
    \begin{subfigure}[b]{0.49\linewidth}
        \centering
        \includegraphics[width=\linewidth]{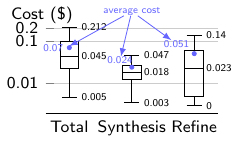}
        \caption{Monetary Cost.}
        \label{fig:eval:monetary_cost}
    \end{subfigure}
    \caption{Time and Monetary Cost on \dataset.}
    \Description{}
    \label{fig:eval:cost}
\end{figure}

\smalltitle{Time and Monetary Cost.}
Beyond effectiveness, we evaluate \sysname’s efficiency in terms of execution time and monetary cost. \autoref{fig:eval:cost} presents Tukey-based boxplots of both metrics across \dataset, decomposed by processing phase. Overall, \sysname completes \poc generation in 255.41s on average, with a mean monetary cost of \$0.07 per case.

To better understand runtime behavior, we decompose total execution time into \textbf{Trace Lifting}, which converts raw traces into structured \poc sketches, and \textbf{PoC Synthesis}, which involves iterative LLM interactions and \work{Foundry}-based validation.
During Trace Lifting, the median runtime is 2.90s, the mean is 67.36s, and the maximum is 600s due to timeout. Most cases (93.46\%) complete within the limit, and 82.87\% finish under 60s. For non-timeout cases (average runtime: 30s), instruction-level lifting and structural optimizations in the dual decompiler account for 22.86s on average, representing the majority of phase runtime.
PoC Synthesis exhibits higher latency, with a median of 103.91s and mean 188.05s. The majority of runtime in this phase arises from network-dependent LLM inference, iteration, and on-chain state retrieval.

Monetary cost is decomposed into \textbf{Synthesis} (mean \$0.024) and \textbf{Refine} (\$0.051), totaling \$0.07 per case (\autoref{fig:eval:monetary_cost}). Synthesis incurs a largely stable cost, whereas Refine exhibits higher variability, reflecting differences in refinement effort, and the minimum Refine cost is \$0, indicating that some cases require no refinement.
Further analysis of LLM token consumption and the number of refinement iterations  reveals that most cases require a single syntactic correction, and some additionally require semantic refinements. Moreover, higher Refine costs are associated with increased token usage and repeated semantic corrections, especially for unresolved semantic inconsistencies could cause more monetary overhead in challenging cases.

\smalltitle{LoC.}
Finally, beyond efficiency and effectiveness, we report a simple property of the generated PoCs as a complementary signal of understandability. On average, the PoCs contain 132 lines of code (median: 116; max: 588), suggesting that they remain human-scale rather than overly verbose artifacts.

\subsection{Failure Analysis} \label{sec:eval:failure_analysis}
To understand the limitations of \sysname, we analyze the failed cases across three dimensions: static analysis timeouts, unresolved syntactic issues, and semantic refinement failures (see~\autoref{fig:eval:main_result}).

\smalltitle{Timeout Analysis.}
21 cases (6.54\%) timed out, occurring during either the trace processor (\S~\ref{sec:designs:tracer}) or the trace-driven decompiler (\S~\ref{sec:designs:dual_decompiler}) stage.
Despite occurring at different stages, all these cases share a common root cause: extreme execution complexity.

Specifically, one timeout occurs during trace processing, where the transaction\footnote{The attack transaction is \href{https://app.blocksec.com/explorer/tx/bsc/0xc7927a68464ebab1c0b1af58a5466da88f09ba9b30e6c255b46b1bc2e7d1bf09}{0xc7927a...}} cannot be fully traced due to extreme execution complexity.
The remaining 20 timeout cases arise from excessive instruction volume in attack functions.
For these cases, the longest functions identified at the locator stage (\S~\ref{sec:designs:locator}) contain on average $\sim$262K instructions (median: 228K), which is $13\times$ (mean) and $71\times$ (median) larger than those in non-timeout cases.
Further inspection reveals that such instruction explosion is typically driven by highly iterative execution patterns.
In 10 cases, loops execute more than 1{,}000 iterations.
For example, in the FIL314 attack\footnote{The attack transaction is \href{https://bscscan.com/tx/0x9f2eb13417190e5139d57821422fc99bced025f24452a8b31f7d68133c9b0a6c}{0x9f23b1...}, and the PoC is available at \href{https://github.com/SunWeb3Sec/DeFiHackLabs/blob/main/src/test/2024-04/FIL314_exp.sol}{FIL314\_exp}.}, execution involves 6{,}000 loop iterations, resulting in 2.776M instructions, with a single function contributing about 740K instructions.

\begin{figure}[t]
\begin{subfigure}{0.46\linewidth}
\centering
\begin{minipage}[b]{\linewidth}
\centering
\begin{lstlisting}[style=solidity_style, ]
call 0x81917e....flashLoan(..., 
  data=abi.encode(
    0x14172fcd41..., /*!\label{zero_omits_entry}!*/
    0x944490e6cb..., /*!\label{zero_omits_exit}!*/
  ) 
)  
\end{lstlisting}
\end{minipage}
\caption{prefix ``0'' are omitted in constant values (Lines 3-4).}
\label{fig:zerp_omitted_in_constant}
\end{subfigure}
\begin{subfigure}{0.52\linewidth}
\centering
\begin{minipage}[b]{\linewidth}
\begin{lstlisting}[style=solidity_style]
bytes memory dpp2Data=
  hex"14172fcd41..."
  hex"944490e6cb..." // error: Expected even number of hex-nibbles./*!\label{zero_omits_error}!*/
  ...;
IDPPAdvanced.flashLoan(.., dpp2Data);
\end{lstlisting}
\end{minipage}
\caption{Synthesized code but with compilation errors (Line 2).}
\label{fig:zero_omits_poc}
\end{subfigure}
\caption{Illustration of syntax errors in synthesized code (the transaction of this DualPool exploit is \href{https://bscscan.com/tx/0x90f374ca33fbd5aaa0d01f5fcf5dee4c7af49a98dc56b47459d8b7ad52ef1e93}{90f374...}).}
\label{fig:syntax_errors}
\Description{}
\end{figure}

\smalltitle{Unresolved Syntax Errors.}
We find that the LLM can iteratively resolve most (93\%) syntax errors (\eg, type mismatches or undeclared variables) through syntax refinement (\S~\ref{sec:designs:refinement}), consistent with prior observations~\cite{disco}.
However, for those unrepaired syntax errors,  $>$60\% of them are attributed to parameter encoding issues.
As shown in \autoref{fig:zerp_omitted_in_constant}, decompilers often omit leading zeros from constants (Lines~\ref{zero_omits_entry}-\ref{zero_omits_exit}).
This results in hexadecimal literals with an \textit{odd} number of digits in the generated \poc (\autoref{fig:zero_omits_poc}), which triggers a Solidity syntax error (Line~\ref{zero_omits_error}).
In the absence of original ABI information, the LLM lacks sufficient semantic context to determine the intended padding and thus fails to repair the malformed literal.

This issue stems from the closed-source nature of most attack contracts.
Without access to ABIs, decompilers (\eg, \work{Panoramix}~\cite{Panoramix}) rely on heuristics to decode parameters, often producing raw constants with ambiguous semantics.
The LLM subsequently propagates these constants verbatim into the generated \poc.
A potential mitigation is to incorporate semantic-aware parameter analysis to help reconstruct the intended argument structure, thereby reducing encoding-related syntax failures.

\smalltitle{Unresolved Semantic Errors.}
As shown in \autoref{fig:eval:main_result}, unresolved semantic errors arise from three distinct failure modes: (1) incomplete context (21.50\%); (2)  execution failures in \work{Foundry} (7.17\%); and (3) limitations in LLM-driven semantic refinement (7.17\%).

\textit{(1) Incomplete Context.}
This failure mode occurs when intelligence sources omit preparatory transactions executed by the attacker prior to the exploit.
In the absence of these state-setting steps, synthesized \poc are executed under invalid preconditions.
We provide detailed examples, including missing allowance initialization in the \texttt{MEEKOO} contract, along with corresponding traces and PoC corrections.
A potential mitigation is to track relevant contract state variables during trace lifting and pre-initialize them in the \work{Foundry} test harness prior to execution.

\textit{(2) Foundry Execution Failures.}
A subset of semantic failures is caused by instability in Remote Procedure Call (RPC) services.
When the RPC provider fails to retrieve on-chain state or returns database errors, \work{Foundry} is unable to simulate the transaction.
Since the semantic refinement loop critically depends on successful execution feedback, such infrastructure-level failures interrupt the debugging process and prevent the synthesis of a verifiable \poc.

\textit{(3) LLM Refinement Failures.}
These failures stem from several complementary challenges.
\textit{First}, execution traces generated by \work{Foundry} often diverge from the original exploit traces in terms of contract addresses and internal call structures, which hinders accurate semantic diffing and root-cause localization.
\textit{Second}, certain semantic defects reflect inherent limitations of current LLMs that cannot be resolved through simple prompting.
For example, in the \texttt{EXcommunity} attack\footnote{The attack transaction is \href{https://bscscan.com/tx/0x5446bf2b57749abdab01813a50ce36246177f3437599f3a56bc1554f596b2c3a}{0x5446bf...}, and the \poc is available at \href{https://github.com/SunWeb3Sec/DeFiHackLabs/blob/main/src/test/2024-05/EXcommunity_exp.sol}{EXcommunity\_exp}.}, the LLM incorrectly interprets a custom \texttt{send} implementation as Solidity’s built-in Ether transfer primitive, leading to flawed logic.
\textit{Third}, deliberate obfuscation in attack contracts (\eg, computing values via \texttt{sha3} over contract addresses) causes synthesized addresses to deviate from the originals, preventing exact reproduction of address-dependent behavior.
These limitations point to the need for future research on robust state reconstruction, trace alignment, and semantics-aware code synthesis.

\subsection{Ablation Studies} \label{sec:eval:ablation_studies}

We conduct a stepwise ablation study to quantify the contribution of each core component in \sysname. To balance evaluation depth with computational cost, we employ a two-stage assessment framework.

The \textit{first} stage evaluates static analysis by measuring the quality of the intermediate \poc (\ie, the \poc sketch) in code structure and redundancy. 
This stage includes two variants:

\begin{packeditemize}
    \item \work{TracExp-NoConcrete}: omits the use of concrete values, evaluating the impact of concrete values on trace-analysis success rate (\ie, whether or not sketch generation) and efficiency.
    \item \work{TracExp-NoCompressor}: omits the compressor, evaluating if trace compression effectively summarizes loops, thereby reducing the token volume and improving human readability.
\end{packeditemize}

The \textit{second} stage assesses \poc generation, measuring the \textit{verifiable} ability and demonstrating the contribution of syntax and semantic refinement. This stage also contains two variants:

\begin{packeditemize}
    \item \work{TracExp-NoSyn}: omits syntactic refinement.
    \item \work{TracExp-NoSem}: omits semantic refinement.
\end{packeditemize}

\begin{figure}[t]
    \centering
    \begin{subfigure}[b]{0.3\linewidth}
        \centering
        \includegraphics[height=2cm]{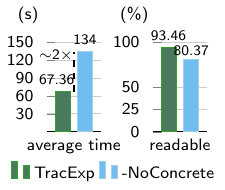}
        \caption{\sysname vs \work{-NoConcrete}}
        \label{fig:eval:ablation_studies_1}
    \end{subfigure}
    \begin{subfigure}[b]{0.33\linewidth}
        \centering
        \includegraphics[height=2cm]{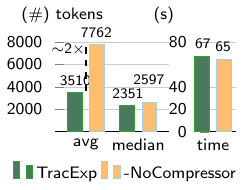}
        \caption{\sysname vs \work{-NoCompressor}}
        \label{fig:eval:ablation_studies_2}
    \end{subfigure}
    \begin{subfigure}[b]{0.33\linewidth}
        \centering
        \includegraphics[height=2cm]{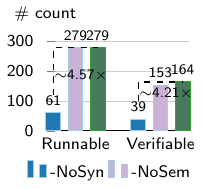}
        \caption{\sysname vs \work{-NoSyn} \& \work{-NoSem}}
        \label{fig:eval:ablation_studies_3}
    \end{subfigure}
    \caption{Ablation studies of \sysname on \dataset.}
    \Description{}
    \label{fig:eval:ablation_studies}
    \vspace{-5mm}
\end{figure}

\autoref{fig:eval:ablation_studies} summarizes the ablation results, confirming that each design component contributes to the efficiency or effectiveness of \poc generation.
\autoref{fig:eval:ablation_studies_1} shows that incorporating concrete-value analysis substantially improves efficiency: the average end-to-end runtime is \textbf{halved}, and the \poc generation success rate increases by $\sim$13\%. \autoref{fig:eval:ablation_studies_2} demonstrates that the compressor algorithm effectively reduces the total token number provided to LLMs, decreasing the average input token volume by $\sim$55\%. The median values are similar because we find that $<$20\% of attacks contain heavy loops. However, for those loop-intensive cases the compressor can reduce the total token count dramatically (in extreme cases to about 1/100 of the original). The additional runtime overhead introduced by compression is negligible ($<$3 seconds on average).

For iterative refinement, \autoref{fig:eval:ablation_studies_3} reports the impact of syntactic and semantic refinement. Without syntax refinement, only a small fraction of generated PoCs are immediately runnable or verifiable ($\sim$20\% and 12.15\%, respectively). Enabling syntax refinement increases the runnable and verifiable rates by factors of 4.57$\times$ and 3.92$\times$, respectively. Adding semantic refinement yields a further improvement: verifiableness increases by an additional 3.43\%.

\subsection{Comparison with Alternative Approaches} \label{sec:eval:comparison}
In this section, we report the comparison results between \sysname and existing exploit generation approaches. To our best of our knowledge, we are the \textit{first} to directly generate \poc code from raw transactions. Thus, we selected alternative approaches that generate exploits from \textit{vulnerable contracts} and \textit{attack contracts}:

\begin{packeditemize}
  \item \work{A1}~\cite{a1}: It proposes an agentic system that transforms LLMs into smart contract exploit generators by equipping them with 6 domain-specific tools (\eg, a source-code fetcher) for analyzing \textit{vulnerable} smart contracts. Here, we exclude \work{AdvScanner}~\cite{AdvSCanner} as it is not open-source and focuses exclusively on reentrancy vulnerabilities. We also exclude \work{Foray}~\cite{foray} since it requires substantial manual effort (\ie, specifying attack targets, initial-state configurations, and additional function mappings) to generate exploits, which limits its scalability on general attacks. 
  \item \work{DiSCo}~\cite{disco}: It translates EVM bytecode into a natural language intermediate representation (IR) and then leverages LLMs to lift this IR into source code, which facilitates attack reproduction. As some of the lifted source codes are not directly executable, we manually adapted them into \work{Foundry}-compatible code and manually validated whether they can reproduce the attacks. Here, we did not include other EVM decompilers (\eg, \work{Gigahorse}~\cite{gigahorse} and \work{Panoramix}~\cite{Panoramix}) because their outputs are often complex \textit{pseudocode} and typically omit logic in contract constructors~\cite{disco}. 
\end{packeditemize}

\smalltitle{Benchmarks.}
For a fair comparison, we evaluate \sysname on the dataset curated by \work{A1}~\cite{a1}, which comprises 38 real-world exploits on Ethereum and BSC (2021--2025). 
This dataset includes 29 incidents from the \work{Verite} benchmark~\cite{verite} and 9 cases introduced by \work{A1}.
Notably, while \work{A1} excluded two cases due to unavailable source code, we include them to demonstrate \sysname's independence from source-level information. We utilize the corresponding \poc codes from \work{DeFiHackLabs}~\cite{DeFiHackLabs} as the ground truth for validation.

\smalltitle{Overall Results.}
Overall, \sysname produced 30 verifiable PoCs (78.95\%) with a monetary cost of \$0.0426 and a time cost of 142.75 seconds on average, representing relative improvements over the baselines ranging from 10\% to 68\%. Across these incidents, \sysname generated PoC code for 95\% of cases, and nearly 90\% of the generated PoCs compiled successfully. These results indicate that \sysname substantially improves both the coverage of automatic PoC synthesis and the rate of producing compilable, executable exploits compared to prior approaches. Additionally, we observe that \sysname performs substantially better on this dataset than on \dataset. Further investigation attributes this gap to differences in the complexity distribution of attack transactions across the two datasets. In particular, the incidents in this dataset require materially less analysis effort: their average runtime and monetary cost are only approximately 60\% of those measured for \dataset, which explains the improved performance.

\smalltitle{Compared with \work{A1}~\cite{a1}.}
Since \work{A1} is not open-source, we can only compare with its reported results. Beyond the final \poc generation result, we also compare the monetary cost. \work{A1} relies on six different models, so for a fair comparison, we consider Gemini Pro with equivalent cost to GPT-5 (\$1.5/M for input, \$10/M for output). The average cost per case for \work{A1} ranges from \$0.11 to \$0.19, while \sysname only incurs an average of \$0.0426, which is significantly lower. One contributing factor is that, for many cases (25 out of 38), the generated code already reproduces the attack effectively immediately after syntax repair, and 7 cases required no syntax refinement at all.
In contrast, the inputs for \work{A1} include long contract code, and it relies on interactions among multiple agents, which inevitably leads to additional token consumption.

\smalltitle{Compared with \work{DiSCo}~\cite{disco}.}
Although \work{DiSCo} can decompile EVM bytecode into source code, our experiments show that only 16 out of 38 attack contracts were successfully decompiled, while the others failed. This observation suggests that traditional decompilation tools may face efficiency issues when reversing complicated attack contracts. In contrast, \sysname analyzes only the transaction traces and also introduces concrete values, which may help alleviate this problem. Moreover, among the successfully decompiled contracts, manual inspection revealed that some code, particularly attack-related logic, was still missing. This finding aligns with \work{DiSCo}'s claim that LLMs may omit content they fail to understand.

\section{Discussion}\label{sec:discussion}

\smalltitle{Data Leakage Concerns.}
Data leakage is a common concern when applying LLMs to code generation. We mitigate this risk through controlled inputs and empirical validation: the model only receives intermediate representations (\ie, decompiled pseudocode and context-rich \poc drafts), and our evaluation includes many attacks without public \poc, with comparable performance across both settings.
Following prior work~\cite{huang2023empirical,disco}, we further assess potential leakage by ranking public PoCs by semantic similarity and manually inspecting 50 pairs of generated and public PoCs. We observe clear stylistic and structural differences and find no exact duplication, suggesting data leakage does not affect our conclusions.

\smalltitle{Manual Efforts for Unverifiable PoCs.}
Even some generated PoCs are not immediately verifiable, \sysname substantially reduces manual effort by producing complete PoCs that captures the correct exploit logic. For PoCs that fail verification, manual inspection confirms consistency with ground-truth attack logic, indicating implementation rather than reasoning errors.
Further, we sample 10 PoCs with syntactic errors and 10 with semantic errors for manual repair. Syntactic issues are resolved within $\sim$5 minutes on average, while semantic issues require $\sim$20 minutes, except for \work{Foundry} execution failures, all sampled cases can be repaired. Overall, \sysname shifts exploit reproduction from manual reverse engineering to lightweight debugging and refinement.

\smalltitle{Attack Coverage.}
We evaluate \sysname across diverse attack patterns (\S~\ref{sec:background}) and observe consistent performance. Across all patterns, \sysname reproduces 49.17$\pm$4\% of attacks, suggesting that it does not rely on pattern-specific heuristics.
Notably, \sysname remains effective for attacks \textit{without} adversary contracts, successfully reproducing 64.29\% of such events. This result indicates that the approach generalizes beyond adversary-contract–centric attack models, provided sufficient execution traces are available.

\smalltitle{Applicability to Other LLMs.}
\sysname is designed to be LLM-agnostic and interacts with models through standard prompt-based interfaces. To validate this design, we further evaluate DeepSeek-R1~\cite{DeepSeek-r1} and Gemini-2.5-Flash~\cite{gemini_25} on \work{A1} datasets (\S~\ref{sec:eval:comparison}). All these LLMs can generate PoCs, though the fraction of verifiable PoCs decreases by $\sim$25\% and 10\%, respectively, reflecting differences in code generation and reasoning capabilities. 
Importantly, the pipeline remains functional across LLMs, and future improvements in base LLMs can directly improve end-to-end reproduction rates.

\smalltitle{Limitations.}
This work has two main limitations: (1) when attack traces contain extremely long execution paths, static analysis may become slow or time out. This stems from our instruction-level lifting of concrete execution paths prior to loop abstraction. While CFG-first decompilation approaches~\cite{Panoramix} scale better, they may lose dynamic flow information critical for accurate reconstruction. A hybrid approach that selectively preserves dynamic information while leveraging CFG-based analysis could improve scalability without sacrificing fidelity.
(2) PoC synthesis currently assumes that users can provide a complete set of attack transactions. Missing preparatory transactions or state updates may cause execution failures, and such state is sometimes difficult to recover in practice. Future work could mitigate this limitation by automatically reconstructing and initializing relevant contract state using inferred read/write dependencies from traces and on-chain state.

\section{Related Work}\label{sec:relatedwork}

In this section, we position \sysname in the context of execution trace analysis, automated exploit generation, bytecode analysis, and the emerging paradigm of LLM-based security analysis.

\smalltitle{Execution Trace Analysis.}  
Execution trace analysis has been widely studied for attack detection. Early systems such as \work{Sereum}~\cite{sereum} and \work{SODA}~\cite{soda} analyze traces using dynamic tainting and instruction-level pattern matching. To capture more complex attack logic, later approaches adopt graph-based representations, including \work{TXSpector}~\cite{txspector}, which constructs Execution Flow Graphs for rule-based reasoning, and \work{Clue}~\cite{clue}, which integrates call, control, and data dependencies into an Execution Property Graph. Recent work further incorporates graph learning or pre-trained models to improve detection accuracy~\cite{evil_under_the_sun, defiguard, PreTS}, with specialized systems targeting price manipulation~\cite{defiranger, defort}, flash loan exploits~\cite{FlashSyn}, and cross-chain attacks~\cite{BridgeGuard}.
While \sysname shares the use of execution traces to capture inter-contract dependencies, prior work primarily focuses on detection. In contrast, \sysname lifts concrete execution traces into human-readable code, bridging detection and verification through automated \poc synthesis.

\smalltitle{Exploits Generation.}
To improve interpretability, several studies generate exploit transactions~\cite{smartest,FlashSyn} or contracts~\cite{foray,cpmmx,AdvSCanner,a1}. \work{Smartest}~\cite{smartest} combines symbolic execution with LLMs to prioritize vulnerable transaction sequences, while \work{Foray}~\cite{foray} and \work{AdvScanner}~\cite{AdvSCanner} generate exploit code using static analysis and LLMs but require substantial manual refinement or target specific vulnerability classes. More recent agentic frameworks, such as \work{A1}~\cite{a1}, \work{CVE-GENIE}~\cite{ullah2025cve}, and \work{SmartPoC}~\cite{smartpoc}, enable autonomous exploit generation using execution-driven agents.
Unlike these approaches, \sysname starts from confirmed attack transactions and directly reconstructs \poc code, enabling general exploit reproduction without relying on prior vulnerability detection or access to source code.

\smalltitle{EVM Bytecode Analysis.}
Numerous EVM decompilers target pseudocode~\cite{gigahorse,elipmoc,Shrnkr,Heimdall,ethervm,erays,Panoramix} or high-level source code~\cite{disco,evm_decompiler}. Pseudocode-oriented tools often preserve low-level semantics at the cost of readability, while source-level decompilers improve readability but may sacrifice semantic fidelity. Existing tools also struggle with obfuscation~\cite{ObfProbe,skanf}, dynamic data types~\cite{BlockWatchdog,smartcat}, and typically aim to recover entire contracts rather than specific logic fragments.
Rather than decompiling bytecode directly, \sysname reverses concrete execution traces, enabling concise and semantically accurate reconstruction of attack logic for \poc generation.

\smalltitle{LLM-based Security Analysis.}
LLMs have been increasingly applied to smart contract security tasks, including logic vulnerability detection~\cite{david2023needmanualsmartcontract,sun2024gptscan,yu2025smart}, DeFi price manipulation analysis~\cite{zhong2025defiscope}, secure program partitioning~\cite{liu2025towards}, and similarity-based vulnerability identification~\cite{zhang2024combining}. Beyond smart contracts, LLMs have also been explored in broader cybersecurity settings, such as automated attack generation~\cite{xu2024autoattacker}, enhanced decompilation~\cite{xu2023lmpa,Tan_2024}, repository-level auditing~\cite{guo2025repoaudit}, and code review generation~\cite{jaoua2025combining}. These efforts primarily leverage LLMs to improve vulnerability discovery, summarization, or reasoning over security-relevant artifacts.
In contrast, \sysname employs LLMs for post-incident exploit reconstruction. Rather than detecting vulnerabilities or ranking alerts, we lift concrete execution into human-readable \poc, complementing prior LLM-based security analysis with a verification-oriented perspective.

\section{Conclusion}\label{sec:conclusion}
We presented \sysname, the \textit{first} automated framework for synthesizing \poc directly from on-chain attack executions. By bridging low-level execution traces with high-level exploit logic, \sysname enables systematic and reproducible understanding of real-world DeFi attacks without requiring source code or prior vulnerability knowledge.
At its core, \sysname combines trace-driven reverse engineering with LLMs to distill concise, semantically faithful exploit logic, and validates synthesized PoCs based on exploitability-relevant semantics. 
Our evaluation on nearly 300 real-world attacks shows that \sysname can automatically generate over 50\% verifiable PoCs for a large fraction of incidents.
By lowering the barrier to post-attack analysis and enabling rapid attack reproduction at scale, \sysname supports more effective incident response and contributes to improving the security of the DeFi ecosystem.

\bibliographystyle{ACM-Reference-Format}

\bibliography{ref.bib}

\end{document}